\documentclass[a4paper,twoside,10pt]{letter}
\usepackage{graphicx,saj,multicol,subeqnarray}
\usepackage{ wasysym }
\usepackage{longtable}
\usepackage{ gensymb }
\usepackage{supertabular}
\usepackage{longtable}
\usepackage{tabularx}
\usepackage{rotating}
\usepackage{pdflscape}

\newcommand{\HII}{H\,{\sc ii}}

\newcommand{\M}{M\,31}

\def\degr{\hbox{$^\circ$}}
\def\arcmin{\hbox{$^\prime$}}
\def\arcsec{\hbox{$^{\prime\prime}$}}

\def\p0{\phantom{0}}

\def\papertype{}

\def\udc{...}
\begin{document}
\baselineskip=3.1truemm
\columnsep=.5truecm
\newenvironment{lefteqnarray}{\arraycolsep=0pt\begin{eqnarray}}
{\end{eqnarray}\protect\aftergroup\ignorespaces}
\newenvironment{lefteqnarray*}{\arraycolsep=0pt\begin{eqnarray*}}
{\end{eqnarray*}\protect\aftergroup\ignorespaces}
\newenvironment{leftsubeqnarray}{\arraycolsep=0pt\begin{subeqnarray}}
{\end{subeqnarray}\protect\aftergroup\ignorespaces}
%


\markboth{\eightrm  20~cm VLA Radio-Continuum Study of {\M\ }-- Images and Point Source Catalogues }
{\eightrm T. J. Galvin et al.}

{\ }

\publ

\type

{\ }


\title{20~cm VLA Radio-Continuum Study of {\M\ }-- Images and Point Source Catalogues }


\authors{T.~J. Galvin, M.~D. Filipovi\'c, E.~J. Crawford,  N.~F.~H. Tothill, G.~F. Wong,  A.~Y. De Horta}

\vskip3mm


\address{University of Western Sydney, Locked Bag 1797, Penrith South DC, NSW 2751, AUSTRALIA}
\Email{m.filipovic}{uws.edu.au}


\dates{April 23, 2012}{April 27, 2012}

\summary{We present a series of new high-sensitivity and high-resolution radio-continuum images of \M\ at $\lambda$=20~cm ($\nu$=1.4~GHz). These new images were produced by merging archived 20~cm radio-continuum observations from the Very Large Array (VLA) telescope. Images presented here are sensitive to rms=60~$\mu$Jy and feature high angular resolution ($<$10\arcsec). A complete sample of discrete radio sources have been catalogued and analysed across 17 individual VLA projects. We identified a total of 864 unique discrete radio sources across the field of \M. One of the most prominent regions in \M\ is the ring feature for which we estimated total integrated flux of 706~mJy at $\lambda$=20~cm. We compare here, detected sources to those listed in Gelfand et al. (2004) at $\lambda$=92~cm and find 118 sources in common to both surveys. The majority (61\%) of these sources exhibit a spectral index of $\alpha <$--0.6 indicating that their emission is predominantly non-thermal in nature. That is more typical for background objects.} 


\begin{multicols}{2}
{

\section{1. INTRODUCTION}

A member of the Andromeda constellation \M\, at a distance of $\sim$778~Kpc (Karachentsev et al. 2004), is the closest spiral galaxy to our own. For this reason, it plays a significant role in galactic and extragalactic studies. A number of previous radio-continuum studies at $\lambda$=20~cm (Braun 1990a) focused on general properties of \M, such as its structure and magnetic fields. Also, Braun (1990b) presented a list of 20-cm sources (534) in the northeast parts of \M. A number of other studies such as Dickel et al. (1982) estimated flux densities of \M\ supernova remnants (SNRs) and \HII\ regions.

In this paper, we reexamine all available archived radio-continuum observations performed with the Very Large Array (VLA) at $\lambda$=20~cm ($\nu$=1.4~GHz) with the intention of merging these observations into a single mosaic radio-continuum image. By combining a large amount of existing data, while using the latest generation of computing power, we can create new images that feature both high angular resolution and improved sensitivity. The newly constructed images are analysed and the difference between the various \M\ images created at 20~cm are discussed. 

In \S2 we describe the observational data and reduction techniques. In \S3 we present our new maps and a brief discussion. Source catalogues are given in \S4 and \S5 is the conclusion.

\section{2. DATA}

A collection of existing, archived radio-continuum observations at $\lambda$=20 cm with pointings centred on \M\ were obtained from the National Radio Astronomy Observatory (NRAO)\footnote{https://archive.nrao.edu/archive/e2earchivex.jsp} online data retrieval system. In total, 15 VLA projects with a variety of array configurations were selected for use in this study, as summarised in Table~1. These projects were observed between the 1$^{st}$ of October 1983 and 27$^{th}$ of September 1997 and are comprised of 28 individual observational runs.

\section{3. Image Creation}

The \textsc{miriad} (Sault et al. 1995) and \textsc{karma} (Gooch 1996) software packages were used for data reduction and analysis. Because of the large volume of data, the \textsc{miriad} package was compiled to run on a 16-processor high-performance computer system. 

Initially, observations were imported into \textsc{aips} using the task \textsc{fillm}, and then all sources were split with \textsc{split}. Using the task \textsc{uvfix}, source coordinates were converted from the B1950 to the J2000 reference frame and the task \textsc{fittp} was used to export each source to a \textsc{fits} file. 

The \textsc{miriad} package was then used for actual data reduction. The task \textsc{atlod} was used to convert ATCA observations into \textsc{miriad} files, while the task \textsc{fits} was used to import the previous \textsc{aips}-produced fits files and convert them to \textsc{miriad} files. Typical calibration and flagging procedures were then carried out (Sault et al. 1995). Using the task \textsc{invert}, each project was imaged individually using a natural weighting scheme. Images with a single pointing were cleaned using the task \textsc{clean}, while each mosaic image was cleaned using the task \textsc{mossdi}. Each of these cleaning tasks uses a SDI clean algorithm to reduce image artefacts (Steer et al. 1984). To convolve a clean model the task \textsc{restor} was then used on each of the cleaned maps, followed by \textsc{linmos} to correct for the primary beam for single pointing observations. For more information on data analysis and image creation see Galvin et al. (2012) and Payne et al. (2004).

The catalogue of radio-continuum sources contains positions RA(J2000), Dec(J2000) and integrated flux densities at 13~cm (Table~A1), 20~cm (Table~A2) and 36~cm (Table~A3).  Table~3 contains the r.m.s., number of sources detected, number of sources identified within the field of the 13~cm image and beam size for each image.   

\section{4. RESULTS}

By comparing the individual maps produced from a variety of observations, the effects of varying array configurations can be seen, as shown in Figs.~1-~18. For example, projects AC0101 and AB0551 show a region of extended emission with poorly resolved point sources across all individual images. This can be attributed to the short baselines of the D type configuration used by the VLA to produce each of these images. 

As expected, observations conducted with a C and B type array configuration, such as AM0464 and AT0149 respectively, demonstrate a progressive loss of extended emission and improved point source resolution across their field of view. Observations that used an A type configuration, such as AH0139 and AH0221, show a significant loss of extended emission but better point source resolution.

}
\end{multicols}

\begin{landscape}
\centerline{{\bf Table 1.}List of VLA projects of \M\ used in this study. RA and DEC represent coordinates of central pointings. }
\vskip1mm
\setlongtables
\small
\begin{longtable}{cccccccccccc}
\hline
$Project$ & $RA$ & $DEC$ & $Array$ & $Date$ & $Freq.$ & $Bandwidth$ & $Time\,on$ & $Recorded$ & $Primary$ & $Secondary$\\
$ID$	& h m s	& \degr\ \arcmin\ \arcsec\	& $Config$ &	& $(MHz)$ & $(MHz)$ & $Source\,(hrs)$&$Polarisation$ & $Calibrator$ & $ Calibrator$ \\
\hline
AH0139 & 00:42:44.24 & 41:16:25.65 & A & 01/10/1983 &1465, 1515&50&1.18& rr,ll,rl,lr & 1328+307 & 0107+562 \\
AH0221 & 00:43:20.37 & 41:13:00.17 & A & 06/04/1986 &1465&50&0.61& rr,ll,rl,lr & 0137+331 & 0026+346 \\
AH0221 & 00:43:44.50 & 41:17:34.87 & A & 06/04/1986 &1465&50&0.61& rr,ll,rl,lr & 0137+331 & 0026+346 \\
AH0221 & 00:40:13.76 & 40:50:06.55 & A & 06/04/1986 &1465&50&0.64& rr,ll,rl,lr & 0137+331 & 0026+346 \\
\smallskip
AH0221 & 00:45:36.88 & 41:03:53.32 & A & 06/04/1986 &1465&50&0.62& rr,ll,rl,lr & 0137+331 & 0026+346 \\
AH0221 & 00:46:00.39 & 42:06:22.99 & A & 06/04/1986 &1465&50&0.56& rr,ll,rl,lr & 0137+331 & 0026+346 \\
AH0221 & 00:46:56.73 & 42:20:27.18 & A & 06/04/1986 &1465&50&0.64& rr,ll,rl,lr & 0137+331 & 0026+346 \\
AH0221 & 00:48:03.74 & 41:40:53.21 & A & 06/04/1986 &1465&50&0.65& rr,ll,rl,lr & 0137+331 & 0026+346 \\
AH0221 & 00:41:30.83 & 40:58:32.59 & A & 06/04/1986 &1465&50&0.49& rr,ll,rl,lr & 0137+331 & 0026+346 \\
\smallskip
AH0221 & 00:42:03.74 & 40:20:31.17 & A & 06/04/1986 &1465&50&0.61& rr,ll,rl,lr & 0137+331 & 0026+346 \\
AH0221 & 00:42:20.03 & 40:57:40.96 & A & 06/04/1986 &1465&50&0.63& rr,ll,rl,lr & 0137+331 & 0026+346 \\
AH0221 & 00:42:30.20 & 41:19:25.83 & A & 06/04/1986 &1465&50&0.63& rr,ll,rl,lr & 0137+331 & 0026+346 \\
AH0221 & 00:42:30.14 & 41:09:25.84 & A & 06/04/1986 &1465&50&0.63& rr,ll,rl,lr & 0137+331 & 0026+346 \\
AH0221 & 00:42:56.31 & 41:19:25.49 & A & 06/04/1986 &1465&50&0.61& rr,ll,rl,lr & 0137+331 & 0026+346 \\
\smallskip
AH0221 & 00:42:56.25 & 41:09:25.50 & A & 06/04/1986 &1465&50&0.61& rr,ll,rl,lr & 0137+331 & 0026+346 \\
AH0221 & 00:43:09.17 & 40:48:03.41 & A & 06/04/1986 &1465&50&0.63& rr,ll,rl,lr & 0137+331 & 0026+346 \\
AT0149 & 00:42:41.02 & 41:15:30.69 & B & 18/04/1993 &1385,1465&50&0.06&rr,ll,rl,lr & 0137+331 & 0248+430 \\
AB0679 & 00:42:46.05 & 41:16:11.63 & C & 26/08/1993 &1365,1465&50&6.55&rr,ll,rl,lr & 0137+331 & 0038+328 \\
AB0679 & 00:42:46:05 & 41:16:11.63 & C & 29/08/1993 &1365,1465&50&14.02& rr,ll,rl,lr & 0137+331 & 0038+328 \\
\smallskip
AH0524 & 00:41:25.89 & 41:12:26.66 & C & 18/12/1994 &1365, 1465&50&4.74& rr,ll,rl,lr & 0137+331 & 0026+346 \\
AM0464 & 00:40:20.00 & 40:31:30.00 & C & 01/11/1994 &1385, 1415&50&1.07& rr,ll,rl,lr & 0137+331 & 0029+349 \\
AM0464 & 00:41:05.00 & 40:46:30.00 & C & 01/11/1994 &1385, 1415&50&1.05& rr,ll,rl,lr & 0137+331 & 0029+349 \\
AM0464 & 00:41:05.00 & 41:04:45.00 & C & 01/11/1994 &1385, 1415&50&1.05& rr,ll,rl,lr & 0137+331 & 0029+349 \\
AM0464 & 00:42:45.00 & 41:04:25.00 & C & 01/11/1994 &1385, 1415&50&1.06& rr,ll,rl,lr & 0137+331 & 0029+349 \\
\smallskip
AM0464 & 00:44:00.00 & 41:34:25.00 & C & 01/11/1994 &1385, 1415&50&1.06& rr,ll,rl,lr & 0137+331 & 0029+349 \\
AM0464 & 00:44:30.00 & 41:22:25.00 & C & 01/11/1994 &1385, 1415&50&1.05& rr,ll,rl,lr & 0137+331 & 0029+349 \\
AM0464 & 00:45:40.00 & 41:47:09.00 & C & 01/11/1994 &1385, 1415&50&1.06& rr,ll,rl,lr & 0137+331 & 0029+349 \\
AC0496 & 00:42:35.44 & 41:57:46.80 & C & 27/09/1997 &1365,1435&50&0.04& rr,ll,rl,lr & 0137+331 & 0102+584 \\
AC0101 & 00:42:44.54 & 41:16:28.65 & D & 13/07/1984 &1465, 1515&50&0.10& rr,ll,rl,lr & 0137+331 & 2234+282 \\
\smallskip
AB0551 & 00:42:46.05 & 41:16:11.63 & D & 27/07/1989 &1465, 1515&50&4.38& rr,ll,rl,lr & 0137+331 & 0038+328 \\
AB0491 & 00:42:05.93 & 40:50:26.15 & D & 08/09/1988 &1465&50&8.66& rr,ll,rl,lr & 1328+307 & 0026+346 \\
AB0647 & 00:44:49.83 & 41:26:23.97 & D & (25,27)/07/1992 &1465,1515&50&3.01, 2.60  & rr,ll,rl,lr & 0137+331 & 0038+328 \\
AB0647 & 00:45:08.07 & 41:51:23.72 & D & (25,27)/07/1992 &1465,1515&50&2.99, 2.57& rr,ll,rl,lr & 0137+331 & 0038+328 \\
AB0437 & 00:43:43.64 & 40:58:27.19 & D & 04/04/1987 &1465,1515&50&8.86& rr,ll,rl,lr & 0137+331 & 0038+328 \\
\smallskip
AB0437 & 00:40:13.44 & 40:46:27.56 & D & 13/04/1987 &1465,1515&50&8.84& rr,ll,rl,lr & 0137+331 & 0038+328 \\
AC0308 & 00:42:00.00 & 40:41:42.00 & D & 08/09/1996 &1365, 1435&50&0.01& rr,ll,rl,lr & 0521+166 & 0141+138 \\
AC0308 & 00:42:00.00 & 41:13:12.00 & D & 08/09/1996 &1365, 1435&50&0.01& rr,ll,rl,lr & 0521+166 & 0141+138 \\
AC0308 & 00:43:30.00 & 40:57:30.00 & D & 08/09/1996 &1365, 1435&50&0.01& rr,ll,rl,lr & 0521+166 & 0141+138 \\
AC0308 & 00:43:30.00 & 41:29:06.00 & D & 08/09/1996 &1365, 1435&50&0.01& rr,ll,rl,lr & 0521+166 & 0141+138 \\
\smallskip
AC0308 & 00:43:30.00 & 42:01:00.00 & D & 08/09/1996 &1365, 1435&50&0.01& rr,ll,rl,lr & 0521+166 & 0141+138 \\
AC0308 & 00:45:00.00 & 41:13:12.00 & D & 20/09/1996 &1365, 1435&50&0.01& rr,ll,rl,lr & 0521+166 & 2202+422 \\
AC0308 & 00:45:00.00 & 42:17:00.00 & D & 20/09/1996 &1365, 1435&50&0.01& rr,ll,rl,lr & 0521+166 & 2202+422 \\
AC0308 & 00:46:30.00 & 41:29:06.00 & D & 20/09/1996 &1365, 1435&50&0.01& rr,ll,rl,lr & 0521+166 & 2202+422 \\
AC0308 & 00:46:30.00 & 42:01:00.00 & D & 20/09/1996 &1365, 1435&50&0.01& rr,ll,rl,lr & 0521+166 & 2202+422 \\
\smallskip
AC0308 & 00:39:00.00 & 40:10:24.00 & D & 24/09/1996 &1365, 1435&50&0.01& rr,ll,rl,lr & 0521+166 & 2202+422 \\
AC0308 & 00:39:00.00 & 40:41:42.00 & D & 24/09/1996 &1365, 1435&50&0.01& rr,ll,rl,lr & 0521+166 & 2202+422 \\
AC0308 & 00:40:30.00 & 40:26:06.00 & D & 24/09/1996 &1365, 1435&50&0.01& rr,ll,rl,lr & 0521+166 & 2202+422 \\
AC0308 & 00:40:30.00 & 40:57:30.00 & D & 24/09/1996 &1365, 1435&50&0.01& rr,ll,rl,lr & 0521+166 & 2202+422 \\
AC0308 & 00:40:30.00 & 41:29:06.00 & D & 24/09/1996 &1365, 1435&50&0.01& rr,ll,rl,lr & 0521+166 & 2202+422 \\
\smallskip
AB0396 & 00:42:46.05 & 41:16:11.62 & B & 27/07/1986 &1465&50&0.97& RCP & 1328+307 & 2352+495 \\
AB0396 & 00:42:46.05 & 41:16:11.62 & B & (18,29)/08/1986 &1465&50&0.94, 0.92& RCP & 1328+307 & 2352+495 \\
AB0396 & 00:42:46.05 & 41:16:11.62 & B & 04/09/1986 &1465&50&1.64& RCP & 1328+307 & 2352+495 \\
AB0396 & 00:43:58.16 & 41:33:32.67 & B & 27/07/1986 &1465&50&0.97& RCP & 1328+307 & 2352+495 \\
AB0396 & 00:43:58.16 & 41:33:32.67 & B & (18,29)/08/1986 &1465&50&0.94, 0.92& RCP & 1328+307 & 2352+495 \\
\smallskip
AB0396 & 00:43:58.16 & 41:33:32.67 & B & 04/09/1986 &1465&50&0.92& RCP & 1328+307 & 2352+495 \\
AB0396 & 00:45:10.98 & 41:50:50.67 & B & 27/07/1986  &1465&50&0.96 &RCP & 1328+307 & 2352+495 \\
AB0396 & 00:45:10.98 & 41:50:50.67 & B & (18,29)/08/1986  &1465&50&0.93, 0.93 &RCP & 1328+307 & 2352+495 \\
AB0396 & 00:45:10.98 & 41:50:50.67 & B & 04/09/1986  &1465&50&0.93 &RCP & 1328+307 & 2352+495 \\
AB0396 & 00:46:24.41 & 42:08:05.64 & B & 27/07/1986 &1465&50&0.80& RCP & 1328+307 & 2352+495 \\
\smallskip
AB0396 & 00:46:24.41 & 42:08:05.64 & B & (18,29)/081986 &1465&50&0.93, 0.92& RCP & 1328+307 & 2352+495 \\
AB0396 & 00:46:24.41 & 42:08:05.64 & B &04/09/1986 &1465&50&0.92& RCP & 1328+307 & 2352+495 \\
AB0396 & 00:44:08.20 & 41:18:06.53 & B & 27/07/1986 &1465&50&0.96& RCP & 1328+307 & 2352+495 \\
AB0396 & 00:44:08.20 & 41:18:06.53 & B & (18,29)/081986 &1465&50&0.92, 0.93& RCP & 1328+307 & 2352+495 \\
AB0396 & 00:44:08.20 & 41:18:06.53 & B & 04/09/1986 &1465&50& 0.91& RCP & 1328+307 & 2352+495 \\
\smallskip
AB0396 & 00:45:20.82 & 41:35:23.54 & B & 27/07/1986 &1465&50&0.95& RCP & 1328+307 & 2352+495 \\
AB0396 & 00:45:20.82 & 41:35:23.54 & B & (18,29)/08/1986 &1465&50&0.92, 0.92& RCP & 1328+307 & 2352+495 \\
AB0396 & 00:45:20.82 & 41:35:23.54 & B & 04/09/1986 &1465&50&0.93& RCP & 1328+307 & 2352+495 \\
AB0396 & 00:46:33.94 & 41:52:38.51 & B & 27/07/1986 &1465&50&0.96& RCP & 1328+307 & 2352+495 \\
AB0396 & 00:46:33.94 & 41:52:38.51 & B & (18,29)/08/1986 &1465&50&0.93, 0.92& RCP & 1328+307 & 2352+495 \\
\smallskip
AB0396 & 00:46:33.94 & 41:52:38.51 & B & 04/09/1986 &1465&50& 0.93& RCP & 1328+307 & 2352+495 \\
AB0396 & 00:42:35.70 & 41:31:37.76 & B & 27/07/1986 &1465&50&0.96& RCP & 1328+307 & 2352+495 \\
AB0396 & 00:42:35.70 & 41:31:37.76 & B & (18,29)/08/1986 &1465&50&0.92, 0.92& RCP & 1328+307 & 2352+495 \\
AB0396 & 00:42:35.70 & 41:31:37.76 & B & 04/09/1986 &1465&50& 0.93& RCP & 1328+307 & 2352+495 \\
AB0396 & 00:43:48.11 & 41:48:58.80 & B & 27/07/1986 &1465&50&0.96& RCP & 1328+307 & 2352+495 \\
\smallskip
AB0396 & 00:43:48.11 & 41:48:58.80 & B & (18,29)/08/1986 &1465&50&0.92, 0.92& RCP & 1328+307 & 2352+495 \\
AB0396 & 00:43:48.11 & 41:48:58.80 & B & 04/09/1986 &1465&50&0.93& RCP & 1328+307 & 2352+495 \\
AB0396 & 00:45:01.14 & 42:06:17.81 & B & 27/07/1986 &1465&50&0.96& RCP & 1328+307 & 2352+495 \\
AB0396 & 00:45:01.14 & 42:06:17.81 & B & (18,29)/08/1986 &1465&50&0.93, 0.93& RCP & 1328+307 & 2352+495 \\
AB0396 & 00:45:01.14 & 42:06:17.81 & B & 04/09/1986 &1465&50&0.93& RCP & 1328+307 & 2352+495 \\
\smallskip
AB0396 & 00:42:46.05 & 41:16:11.62 & C & 14/12/1986 &1465&50&0.93& RCP & 1328+307 & 2352+495 \\
AB0396 & 00:43:58.16 & 41:33:32.67 & C & 14/12/1986 &1465&50&0.93& RCP & 1328+307 & 2352+495 \\
AB0396 & 00:45:10.98 & 41:50:50.67 & C & 14/12/1986 &1465&50&0.92& RCP & 1328+307 & 2352+495 \\
AB0396 & 00:46:24.41 & 42:08:05.64 & C & 14/12/1986 &1465&50&0.92& RCP & 1328+307 & 2352+495 \\
AB0396 & 00:44:08.20 & 41:18:06.53 & C & 14/12/1986 &1465&50&0.93& RCP & 1328+307 & 2352+495 \\
\smallskip
AB0396 & 00:45:20.82 & 41:35:23.54 & C & 14/12/1986 &1465&50&0.94& RCP & 1328+307 & 2352+495 \\
AB0396 & 00:46:33.94 & 41:52:38.51 & C & 14/12/1986 &1465&50&0.93& RCP & 1328+307 & 2352+495 \\
AB0396 & 00:42:35.70 & 41:31:37.76 & C & 14/12/1986 &1465&50&0.93& RCP & 1328+307 & 2352+495 \\
AB0396 & 00:43:48.11 & 41:48:58.80 & C & 14/12/1986 &1465&50&0.93& RCP & 1328+307 & 2352+495 \\
AB0396 & 00:45:01.14 & 42:06:17.81 & C & 14/12/1986 &1465&50&0.89& RCP & 1328+307 & 2352+495 \\
\smallskip
AB0396 & 00:45:10.98 & 41:50:50.67 & D & 13/03/1986 &1465&50&0.90& RCP & 1328+307 & 2352+495 \\
AB0396 & 00:46:24.41 & 42:08:05.64 & D & 13/03/1986 &1465&50&0.85& RCP & 1328+307 & 2352+495 \\
AB0396 & 00:45:20.82 & 41:35:23.54 & D & 13/03/1986 &1465&50&0.74& RCP & 1328+307 & 2352+495 \\
AB0396 & 00:46:33.94 & 41:52:38.51 & D & 13/03/1986 &1465&50&0.80& RCP & 1328+307 & 2352+495 \\
AB0396 & 00:43:48.11 & 41:48:58.80 & D & 13/03/1986 &1465&50&0.89& RCP & 1328+307 & 2352+495 \\
\smallskip
AB0396 & 00:45:01.14 & 42:06:17.81 & D & 13/03/1986 &1465&50&0.88& RCP & 0137+331 & 0137+331 \\
AB0999 & 00:42:46.05 & 41:16:11.62 & D & 23/01/1986 &1465&50&0.20& RCP & 0137+331 & 0137+331 \\
AB0999 & 00:43:58.16 & 41:33:32.67 & D & 23/01/1986 &1465&50&0.26& RCP & 0137+331 & 0137+331 \\
AB0999 & 00:45:10.98 & 41:50:50.67 & D & 23/01/1986 &1465&50&0.20& RCP & 0137+331 & 0137+331 \\
AB0999 & 00:46:24.41 & 42:08:05.64 & D & 23/01/1986 &1465&50&0.17& RCP & 0137+331 & 0137+331 \\
\smallskip
AB0999 & 00:44:08.20 & 41:18:06.53 & D & 23/01/1986 &1465&50&0.27& RCP & 0137+331 & 0137+331 \\
AB0999 & 00:45:20.82 & 41:35:23.54 & D & 23/01/1986 &1465&50&0.27& RCP & 0137+331 & 0137+331 \\
AB0999 & 00:46:33.94 & 41:52:38.51 & D & 23/01/1986 &1465&50&0.25& RCP & 0137+331 & 0137+331 \\
AB0999 & 00:42:35.70 & 41:31:37.76 & D & 23/01/1986 &1465&50&0.27& RCP & 0137+331 & 0137+331 \\
AB0999 & 00:43:48.11 & 41:48:58.80 & D & 23/01/1986 &1465&50&0.27& RCP & 0137+331 & 0137+331 \\
\smallskip
AB0999 & 00:45:01.14 & 42:06:17.81 & D & 23/01/1986 &1465&50&0.27& RCP & 0137+331 & 0137+331 \\
\hline
\end{longtable}
\end{landscape}
\vskip.5cm
\begin{multicols}{2}
{


}
\end{multicols}

\newpage 
\centerline{{\bf Table 2.}The details of VLA single and merged projects of \M\ mosaics at 20~cm. }
\vskip1mm
\centerline{
\small
\begin{tabular}{lcccccc}
\hline
\emph{VLA}&\emph{Beam Size}&\emph{r.m.s.} & Figure\\
\emph{Project}& (arcsec) & (mJy/beam)\\
\hline
AC0101-a  & 45.9$\times$43.2    & 0.49 & 1\\
AB0551-a & 35.9$\times$32.1    & 0.12 & 2\\
AB0491-a  & 39.0$\times$33.8       & 0.12 &3\\
AB0647-a  & 41.2$\times$37.3     & 0.24 & 4\\
\smallskip
AB0647-b  & 40.9$\times$35.4     & 0.24 & 5\\
AB0437-a & 36.0$\times$31.0    & 0.10 & 6\\
AB0437-b  & 36.0$\times$31.1   & 0.19 & 7\\
AC0308-a & 57.9$\times$49.8    &0.72 & 8\\
AC0308-b & 58.9$\times$48.9    &0.54 & 9\\
\smallskip
AC0308-c & 58.0$\times$50.2  & 0.60 & 10 \\
AC0496-a & 54.6$\times$14.0 & 0.08 & 11 \\
AM0464-a & 12.8$\times$12.2 & 0.13 & 12 \\
AH0524-a & 12.8$\times$12.2 & 0.07 & 13 \\
AB0679-a & 12.0$\times$11.7 & 0.07 & 14 \\
\smallskip
AB0679-b & 12.1$\times$11.5 & 0.08 & 15 \\
AT0149-a & 4.0$\times$3.4 & 0.08 & 16 \\
AH0221-a & 3.4$\times$3.2 &0.22 & 17 \\
AH0139-a & 7.2$\times$6.6 & 0.16 & 18 \\
Fully Polarised & 35.73$\times$16.38 & 0.15 & 19 \\
\smallskip
(5 k$\lambda$ restricted) &&& \\
Fully Polarised & 6.4$\times$5.4 & 0.09 & 20 \\
(25 k$\lambda$ restricted) &&& \\
Mosaic AB0396 and AB0999 & 4.6$\times$3.8 & 0.08 & 21 \\
(35 k$\lambda$ restricted) &&& \\
All Calibrated & 32.6$\times$16.4 & 0.13 & 22 \\
(5 k$\lambda$ restricted) &&& \\
All Calibrated  & 6.1$\times$5.4 & 0.12 & 23 \\ 
(25 k$\lambda$ restricted) &&& \\
\hline
\end{tabular}}
\vskip.5cm

\begin{multicols}{2}
{


\subsection{4.1 New Combined \M\ Mosaics at $\lambda$=20~cm}

Figs.~19 and 20 are the resulting images when all fully polarised VLA observations are merged together. Both images suffer from artefacts around the outer region of the field of view. This can be attributed to the image stacking process, where observations conducted with the use of a compact array configuration get stretched to meet the resolution of the image as a whole.  

Fig.~19 shows the resulting radio-continuum image when all fully polarised VLA observations are merged together with a restricted {\it uv} coverage of \mbox{0-5~k$\lambda$}. This restriction was introduced in order to preserve the intricate structure of the extended emission while partly resolving point sources across the field. 

Fig.~20 is the same data-set as Fig.19 with a restricted {\it uv} coverage of 0-25~k$\lambda$. This restriction was imposed after a trial and error process where we identified the \textsc{miriad}'s software limitations. Despite this restriction, point sources are well resolved and there remains a region of extended emission.


Fig.~21 shows a mosaic radio-continuum image of VLA projects AB0396 and AB0999 with a restricted {\it uv} coverage of 0-35~k$\lambda$. Point sources are seen prominently across the field of view with little extended emission. This can be attributed to the larger array configurations, and thus longer baselines, which these observations are constructed of. 

Fig.~22 is a mosaic radio-continuum image comprised of all calibrated VLA observations from this study with a restricted {\it uv} coverage of 0-5~k$\lambda$. This restriction was implemented to place greater emphasis on the intrinsic structure of the extended emission throughout the field of view. The majority of observations within VLA projects AB0396 and AB0999 were made with B configuration types. This provided {\it uv} coverage data that was noticeably absent in other observations. This has significantly improved the overall clarity of the image when compared to Fig.~19. One of the most prominent regions in \M\ is the ring feature for which we estimated total integrated flux of 706$\pm$35~mJy.

In Fig.~23, we show the resulting image when all calibrated VLA observations are merged together with a restricted {\it uv} coverage of 0-25~k$\lambda$. Again, this restricted {\it uv} coverage was used to overcome the limit of the \textsc{miriad} software imaging capabilities. Point sources are well resolved and there remains a region of extended emission.

\section{5. DISCRETE RADIO-CONTINUUM SOURCES IN THE FIELD OF \M } 

For each project imaged, a source catalogue was created. Tables~3,  4, 5, 6, 7, 8, 9, 10 ,11 ,12, 13, 14, 15, 16, 17, 18, 19 and 20
list sources found in each individual project that has been imaged in this study. These catalogues contain a source's RA and DEC positions (J2000) and integrated flux density. All catalogues have been cross referenced and sources common to multiple projects have been noted in Col.~6 of each table. 

Across fifteen individual and three merged projects, a total of 864 unique discrete sources are identified. We compared these discrete sources to those listed in Gelfand et al. (2004) at $\lambda$=92~cm and found 118 sources in common to both surveys. Table~21 is an extract of this comparison, where Col.~11 is the estimated spectral index $(S_\nu \propto \nu^\alpha)$ of each source. The complete list and all catalogues can be found in on-line archive (http://cds.u-strasbg.fr/).

The average flux density, as listed in Col.~5 of Table~21, was calculated by averaging flux density from each project where a discrete source was found. A sources error, as listed in Col.~6, was calculated by finding the largest difference between the average flux density of a source, and the flux density from each project it appeared in. In the case where a source was found in multiple projects, its name, project, RA and DEC, as listed in Cols.~1, 2, 3 and 4, were taken from the highest resolution image. 

In Fig~24 we compare the RA and DEC between our 20~cm catalogue and Gelfand et al. (2004) sources as listed in Table~21. The concentration of points near the centre of graph indicates an accurate model for comparison. We found that the average positional difference in $\Delta$RA and $\Delta$DEC is --0.01\arcsec\ (with a SD of 1.912\arcsec) and +0.18\arcsec\ (with a SD of 1.543\arcsec) respectively. 

Fig~25 shows the spectral index distribution of sources listed in Table~21. The majority (61\%) of sources exhibit a spectral index of $<$--0.6 indicating that their emission is predominantly non-thermal in nature. This implies that most of these sources will be background ANGs or Quasars. Some of these background source could qualify as compact steep spectrum sources.

}
\end{multicols}

\centerline{{\bf Table 3.}  Sample list of sources at 20~cm found in Project AB0437-a}
\vskip1mm
\centerline{
\begin{tabular}{cccccl}
\hline
1   & 2    & 3       & 4       & 5     & 6 \\
 \# & Name & RA      & DEC     & Flux  & Notes \\
    &      & (J2000) & (J2000) & (mJy) &\\
\hline
 1  &  J003904+410822  &  00:39:04.30 &  +41:08:22.20 &  2.46  & \\
 2  &  J003907+410346  &  00:39:07.94 &  +41:03:46.10 &  20.65  & \\
 3  &  J003908+410338  &  00:39:08.28 &  +41:03:38.21 &  7.59  & \\
 4  &  J003918+410301  &  00:39:18.15 &  +41:03:01.10 &  12.43  & \\
 5  &  J003918+411634  &  00:39:18.88 &  +41:16:34.06 &  5.65  & \\
\hline
\end{tabular}}
\vskip.5cm

\centerline{{\bf Table 4.}  Sample list of sources at 20~cm found in Project AB0437-b.}
\vskip1mm
\centerline{
\begin{tabular}{cccccl}
\hline
1   & 2    & 3       & 4       & 5     & 6 \\
 \# & Name & RA      & DEC     & Flux  & Notes \\
    &      & (J2000) & (J2000) & (mJy) &\\
\hline
 1  &  J003839+403300  &  00:38:39.00 &  +40:33:00.60 &  4.85  & \\
 2  &  J003908+403007  &  00:39:08.25 &  +40:30:07.71 &  2.88  & \\
 3  &  J003908+410335  &  00:39:08.39 &  +41:03:35.80 &  6.62  & \\
 4  &  J003917+410258  &  00:39:17.98 &  +41:02:58.90 &  9.85  & \\
 5  &  J003927+405425  &  00:39:27.05 &  +40:54:25.00 &  4.61  & \\
\hline
\end{tabular}}
\vskip.5cm

\centerline{{\bf Table 5.} Sample list of sources at 20~cm found in Project AB0491-a. Column 6 describes the source in additional projects.}
\vskip1mm
\centerline{
\begin{tabular}{cccccl}
\hline
1   & 2    & 3       & 4       & 5     & 6 \\
 \# & Name & RA      & DEC     & Flux  & Notes \\
    &      & (J2000) & (J2000) & (mJy) &\\
\hline
1 & J004013+405005 & 00:40:13.84 & +40:50:05.9 & 41.99 &Table:17 \#6\\\shrinkheight{2em}
2 & J004017+405824 & 00:40:17.03 & +40:58:24.8 & 21.55  &Table:17 \#8\\
3 & J004024+410711 & 00:40:24.78 & +41:07:11.3 & 26.70  &Table:19 \#24\\
4 & J004036+411910 & 00:40:36.91 & +41:19:10.9 & 15270.00  &\\
5 & J004044+404845 & 00:40:44.56 & +40:48:45.3 & 13.30  &\\
\hline
\end{tabular}}
\vskip.5cm

\newpage
\centerline{{\bf Table 6.} Sample list of sources at 20~cm found in Project AB0551-a. Column 6 describes the source in additional projects.}
 \label{table:M31AB0437-b}
\vskip1mm
\centerline{
\begin{tabular}{cccccl}
\hline
1   & 2    & 3       & 4       & 5     & 6 \\
 \# & Name & RA      & DEC     & Flux  & Notes \\
    &      & (J2000) & (J2000) & (mJy) &\\
\hline
1  &  J004056+405734  &  00:40:56.81 &  +40:57:34.31 &  13.90  & \\
2  &  J004100+411358  &  00:41:00.71 &  +41:13:58.30 &  5.53  & \\
3  &  J004109+412456  &  00:41:09.61 &  +41:24:56.80 &  27.57  & \\
4  &  J004117+412316  &  00:41:17.91 &  +41:23:16.22 &  2.16  & \\
5 & J004120+411044 & 00:41:20.12 & +41:10:44.7  & 18.97 &Table:9 \#3, Table:10 \#5\\
\hline
\end{tabular}}
\vskip.5cm

\centerline{{\bf Table 7.} Sample list of sources at 20~cm found in Project AB0647-a.}
\vskip1mm
\centerline{
\begin{tabular}{cccccl}
\hline
1   & 2    & 3       & 4       & 5     & 6 \\
 \# & Name & RA      & DEC     & Flux  & Notes \\
    &      & (J2000) & (J2000) & (mJy) &\\
\hline
 1  &  J004132+412429  &  00:41:32.19 &  +41:24:29.20 &  2.13  & \\
 2  &  J004139+414252  &  00:41:39.59 &  +41:42:52.90 &  2.87  & \\
 3  &  J004139+413040  &  00:41:39.64 &  +41:30:40.80 &  5.69  & \\
 4  &  J004155+413720  &  00:41:55.95 &  +41:37:20.10 &  1.93  & \\
 5  &  J004212+414828  &  00:42:12.82 &  +41:48:28.00 &  4.40  & \\
\hline
\end{tabular}}
\vskip.5cm

\centerline{{\bf Table 8.} Sample list of sources at 20~cm found in Project AB0647-b. Column 6 describes the source in additional projects..}
\vskip1mm
\centerline{
\begin{tabular}{cccccl}
\hline
1   & 2    & 3       & 4       & 5     & 6 \\
 \# & Name & RA      & DEC     & Flux  & Notes \\
    &      & (J2000) & (J2000) & (mJy) &\\
\hline
1 & J004200+415408 & 00:42:00.51 & +41:54:08.4 & 0.76   &\\
2 & J004204+412932 & 00:42:04.53 & +41:29:32.3 & 2.41    &\\
3 & J004218+412930 & 00:42:18.89 & +41:29:30.6 & 155.90  &Table:7 \#6\\
4 & J004233+412929 & 00:42:33.41 & +41:29:29.7 & 2.92    &\\
5 & J004235+415743 & 00:42:35.77 & +41:57:43.4 & 24.40   &Table:7 \#7\\
\hline
\end{tabular}}
\vskip.5cm

\centerline{{\bf Table 9.} Sample list of sources at 20~cm found in Project AB0679-a. Column 6 describes the source in additional projects.}
\vskip1mm
\centerline{
\begin{tabular}{cccccl}
\hline
1   & 2    & 3       & 4       & 5     & 6 \\
 \# & Name & RA      & DEC     & Flux  & Notes \\
    &      & (J2000) & (J2000) & (mJy) &\\
\hline
1 & J004108+412454 & 00:41:08.11 & +41:24:54.7 & 10.76 &Table:10 \#1\\
2 & J004112+412458 & 00:41:12.10 & +41:24:58.6 & 44.01  &\\
3 & J004120+411045 & 00:41:20.18 & +41:10:45.3 & 19.68 &Table:6 \#5, Table:10 \#5\\
4 & J004139+413031 & 00:41:39.60 & +41:30:31.3 & 32.32  &Table:12 \#20, Table:6 \#7, Table:10 \#6\\
5 & J004141+410338 & 00:41:41.50 & +41:03:38.9 & 45.71  &Table:6 \#8, Table:10 \#8\\
\hline
\end{tabular}}
\vskip.5cm

\centerline{{\bf Table 10.} Sample list of sources at 20 cm found in Project AB0679-b. Column 6 describes the source in additional projects. }
\vskip1mm
\centerline{
\begin{tabular}{cccccl}
\hline
1   & 2    & 3       & 4       & 5     & 6 \\
 \# & Name & RA      & DEC     & Flux  & Notes \\
    &      & (J2000) & (J2000) & (mJy) &\\
\hline
 1  &  J004108+412455  &  00:41:08.18 &  +41:24:55.20 &  13.54  & \\
 2  &  J004112+412458  &  00:41:12.32 &  +41:24:58.10 &  4.45  & \\
 3  &  J004114+412454  &  00:41:14.03 &  +41:24:54.40 &  3.03  & \\
 4  &  J004119+412314  &  00:41:19.17 &  +41:23:14.37 &  1.09  & \\
5 & J004120+411045 & 00:41:20.17 & +41:10:45.1 & 19.05  &Table:6 \#5\\
\hline
\end{tabular}}
\vskip.5cm

\newpage
\centerline{{\bf Table 11.} Sample list of sources at 20~cm found in Project AC0101-a. }
\vskip1mm
\centerline{
\begin{tabular}{cccccl}
\hline
1   & 2    & 3       & 4       & 5     & 6 \\
 \# & Name & RA      & DEC     & Flux  & Notes \\
    &      & (J2000) & (J2000) & (mJy) &\\
\hline
 1  &  J004057+412133  &  00:40:57.97 &  +41:21:33.30 &  10.70  & \\
 2  &  J004107+412129  &  00:41:07.74 &  +41:21:29.70 &  6.36  & \\
 3  &  J004108+412444  &  00:41:08.09 &  +41:24:44.50 &  13.49  & \\
 4  &  J004120+411042  &  00:41:20.00 &  +41:10:42.30 &  15.50  & \\
 5  &  J004139+413035  &  00:41:39.65 &  +41:30:35.70 &  27.95  & \\
\hline
\end{tabular}}
\vskip.5cm

\centerline{{\bf Table 12.} Sample list of sources at 20~cm found in Project AC0308-a.}
\vskip1mm
\centerline{
\begin{tabular}{cccccl}
\hline
1   & 2    & 3       & 4       & 5     & 6 \\
 \# & Name & RA      & DEC     & Flux  & Notes \\
    &      & (J2000) & (J2000) & (mJy) &\\
\hline
 1  &  J003938+410327  &  00:39:38.55 &  +41:03:27.90 &  5.31  & \\
 2  &  J003957+411138  &  00:39:57.13 &  +41:11:38.70 &  11.93  & \\
 3  &  J004002+412634  &  00:40:02.34 &  +41:26:34.40 &  5.37  & \\
 4  &  J004010+411825  &  00:40:10.88 &  +41:18:25.90 &  3.47  & \\
 5  &  J004014+410841  &  00:40:14.02 &  +41:08:41.10 &  14.23  & \\
\hline
\end{tabular}}
\vskip.5cm

\centerline{{\bf Table 13.} Sample list of sources at 20 cm found in Project AC0308-b.}
\vskip1mm
\centerline{
\begin{tabular}{cccccl}
\hline
1   & 2    & 3       & 4       & 5     & 6 \\
 \# & Name & RA      & DEC     & Flux  & Notes \\
    &      & (J2000) & (J2000) & (mJy) &\\
\hline
 1  &  J004257+411643  &  00:42:57.84 &  +41:16:43.80 &  8.25  & \\\shrinkheight{2em}
 2  &  J004335+421219  &  00:43:35.23 &  +42:12:19.90 &  3.64  & \\
 3  &  J004337+412019  &  00:43:37.28 &  +41:20:19.40 &  8.92  & \\
 4  &  J004341+405435  &  00:43:41.76 &  +40:54:35.90 &  9.40  & \\
 5  &  J004345+412839  &  00:43:45.22 &  +41:28:39.90 &  4.91  & \\
\hline
\end{tabular}}
\vskip.5cm

\centerline{{\bf Table 14.} Sample list of sources at 20 cm found in Project AC0308-c.}
\vskip1mm
\centerline{
\begin{tabular}{cccccl}
\hline
1   & 2    & 3       & 4       & 5     & 6 \\
 \# & Name & RA      & DEC     & Flux  & Notes \\
    &      & (J2000) & (J2000) & (mJy) &\\
\hline
 1  &  J003651+404452  &  00:36:51.36 &  +40:44:52.40 &  4.81  & \\
 2  &  J003724+403821  &  00:37:24.96 &  +40:38:21.30 &  2.61  & \\
 3  &  J003730+401239  &  00:37:30.21 &  +40:12:39.10 &  4.95  & \\
 4  &  J003745+402513  &  00:37:45.61 &  +40:25:13.50 &  22.56  & \\
 5  &  J003807+405252  &  00:38:07.88 &  +40:52:52.90 &  3.04  & \\
\hline
\end{tabular}}
\vskip.5cm

\centerline{{\bf Table 15.} Sample list of sources at 20 cm found in Project AC0496-a.}
\vskip1mm
\centerline{
\begin{tabular}{cccccl}
\hline
1   & 2    & 3       & 4       & 5     & 6 \\
 \# & Name & RA      & DEC     & Flux  & Notes \\
    &      & (J2000) & (J2000) & (mJy) &\\
\hline
 1  &  J004057+415438  &  00:40:57.96 &  +41:54:38.20 &  1.17  & \\
 2  &  J004105+414451  &  00:41:05.89 &  +41:44:51.43 &  1.28  & \\
 3  &  J004112+415643  &  00:41:12.36 &  +41:56:43.10 &  1.63  & \\
 4  &  J004114+414339  &  00:41:14.21 &  +41:43:39.31 &  1.18  & \\
 5  &  J004129+413536  &  00:41:29.23 &  +41:35:36.41 &  1.02  & \\
\hline
\end{tabular}}
\vskip.5cm

\newpage
\centerline{{\bf Table 16.} Sample list of sources at 20~cm found in Project AH0139-a. Column 6 describes the source in additional projects.}
\vskip1mm
\centerline{
\small
\begin{tabular}{cccccl}
\hline
1   & 2    & 3       & 4       & 5     & 6 \\
 \# & Name & RA      & DEC     & Flux  & Notes \\
    &      & (J2000) & (J2000) & (mJy) &\\
\hline
1 & J004141+410343 & 00:41:41.34 & +41:03:43.0 & 33.45  &\\
2 & J004147+411847 & 00:41:47.93 & +41:18:48.0 & 30.44  &Table:17 \#20, Table:6 \#11, Table:20 \#11,Table:11 \#7, \\
&&&&& Table:9 \#8, Table:10 \#11\\
3 & J004151+411439 & 00:41:51.14 & +41:14:39.4 & 17.68 &Table:6 \#12, Table:9 \#9, Table:10 \#12\\
4 & J004218+412922 & 00:42:18.63 & +41:29:22.6 & 245.11  &\\
5 & J004222+410805 & 00:42:22.03 & +41:08:05.4 & 2.50  &Table:9 \#18, Table:10 \#20\\
\hline
\end{tabular}}
\vskip.5cm

\centerline{{\bf Table 17.} Sample list of sources at 20~cm found in Project AH0221-a. }
\vskip1mm
\centerline{
\begin{tabular}{cccccl}
\hline
1   & 2    & 3       & 4       & 5     & 6 \\
 \# & Name & RA      & DEC     & Flux  & Notes \\
    &      & (J2000) & (J2000) & (mJy) &\\
\hline
 1  &  J003933+404405  &  00:39:33.35 &  +40:44:05.34 &  4.16  & \\
 2  &  J003949+410421  &  00:39:49.41 &  +41:04:21.22 &  6.32  & \\
 3  &  J004012+410840  &  00:40:12.28 &  +41:08:40.73 &  2.85  & \\
 4  &  J004013+410839  &  00:40:13.28 &  +41:08:39.05 &  5.16  & \\
 5  &  J004013+410836  &  00:40:13.69 &  +41:08:36.59 &  6.31  & \\
\hline
\end{tabular}}
\vskip.5cm

\centerline{{\bf Table 18.} Sample list of sources at 20~cm found in Project AH0524-a. Column 6 describes the source in additional projects.}
\vskip1mm
\centerline{
\begin{tabular}{cccccl}
\hline
1   & 2    & 3       & 4       & 5     & 6 \\
 \# & Name & RA      & DEC     & Flux  & Notes \\
    &      & (J2000) & (J2000) & (mJy) &\\
\hline
1 & J003918+410301 & 00:39:18.49 & +41:03:01.0 & 9.27 &Table:19 \#3\\
2 & J003922+411040 & 00:39:22.15 & +41:10:40.9 & 3.70  &\\
3 & J003931+411511 & 00:39:31.41 & +41:15:11.8 & 2.36 &\\
4 & J003932+410440 & 00:39:32.22 & +41:04:40.7 & 1.17 &\\
5 & J003935+411432 & 00:39:35.96 & +41:14:32.7 & 4.74  &Table:14 \#37\\
\hline
\end{tabular}}
\vskip.5cm

\centerline{{\bf Table 19.} Sample list of sources at 20~cm found in Project AM0464-a. }
\vskip1mm
\centerline{
\begin{tabular}{cccccl}
\hline
1   & 2    & 3       & 4       & 5     & 6 \\
 \# & Name & RA      & DEC     & Flux  & Notes \\
    &      & (J2000) & (J2000) & (mJy) &\\
\hline
 1  &  J003908+403009  &  00:39:08.78 &  +40:30:09.99 &  1.84  & \\
 2  &  J003916+403629  &  00:39:16.37 &  +40:36:29.37 &  0.59  & \\
 3  &  J003918+410300  &  00:39:18.55 &  +41:03:00.50 &  5.41  & \\
 4  &  J003919+402206  &  00:39:19.50 &  +40:22:06.08 &  0.74  & \\
 5  &  J003922+411038  &  00:39:22.16 &  +41:10:38.50 &  3.01  & \\
\hline
\end{tabular}}
\vskip.5cm

\centerline{{\bf Table 20.} Sample list of sources at 20~cm found in Project AT0149-a.}
\vskip1mm
\centerline{
\begin{tabular}{cccccl}
\hline
1   & 2    & 3       & 4       & 5     & 6 \\
 \# & Name & RA      & DEC     & Flux  & Notes \\
    &      & (J2000) & (J2000) & (mJy) &\\
\hline
 1  &  J004036+412404  &  00:40:36.79 &  +41:24:04.70 &  3.23  & \\
 2  &  J004037+412051  &  00:40:37.42 &  +41:20:51.49 &  3.78  & \\
 3  &  J004046+411637  &  00:40:46.52 &  +41:16:37.03 &  2.26  & \\
 4  &  J004049+411226  &  00:40:49.05 &  +41:12:26.92 &  2.30  & \\
 5  &  J004054+412632  &  00:40:54.52 &  +41:26:32.00 &  3.88  & \\
\hline
\end{tabular}}
\vskip.5cm


\begin{landscape}
\centerline{{\bf Table 21.} Flux density comparison (sample) between sources in common to $\lambda$=20~cm and $\lambda$=92~cm surveys of the \M. Columns }
\centerline{1, 2, 3 and 4 describes source information from the highest 20-cm resolution project available. Columns 5 and 6  from sources }
\centerline{ common across projects and integrated flux density (Col.~5) represent an average flux density across the are derived various}
\centerline{ project detections. Columns 7, 8, 9 and 10 are from Gelfand et al. (2004). Column 11 is the spectral index of Col.~5 and 9.}

\vskip1mm
\small
\begin{tabular}{cccccccccccc}
\hline
   & 1      & 2       & 3       & 4       & 5           & 6                   & 7       & 8       & 9           & 10 & 11 \\
\# & Source & Project & RA      & DEC     & S$_{20-cm}$ & $\Delta$S$_{20-cm}$ & RA      & DEC     & S$_{92-cm}$ & $\Delta$S$_{92-cm}$ & $\alpha$ \\
   & Name   & \#      & (J2000) & (J2000) & (mJy)       & (mJy)               & (J2000) & (J2000) & (mJy)       & (mJy) & \\
\hline
  1  &  J003807+405252  &  AC0308-c  &  00:38:07.88 &  +40:52:52.90 &  3.04  &  0.06  &  00:38:7.90 &  +40:52:53.75 &  7.03  &  0.67  &  -0.54933 \\
  2  &  J003904+410822  &  AB0437-a  &  00:39:04.30 &  +41:08:22.20 &  2.46  &  0.06  &  00:39:4.34 &  +41:08:18.21 &  8.54  &  0.69  &  -0.81555 \\
  3  &  J003918+411636  &  AC0308-c  &  00:39:18.45 &  +41:16:36.30 &  4.08  &  0.06  &  00:39:18.23 &  +41:16:36.85 &  23.39  &  1.29  &  -1.14424 \\
  4  &  J003927+405425  &  AB0437-b  &  00:39:27.05 &  +40:54:25.00 &  4.49  &  0.12  &  00:39:27.03 &  +40:54:26.57 &  10.48  &  0.74  &  -0.55542 \\
  5  &  J003933+404401  &  AM0464-a  &  00:39:33.02 &  +40:44:01.40 &  9.84  &  2.73  &  00:39:32.79 &  +40:43:59.63 &  18.14  &  2.85  &  -0.40081 \\
  \smallskip
  6  &  J003932+400837  &  AC0308-c  &  00:39:32.83 &  +40:08:37.90 &  50.17  &  0.06  &  00:39:32.82 &  +40:08:36.74 &  126.12  &  8.99  &  -0.60404 \\
  7  &  J003948+403433  &  AM0464-a  &  00:39:48.29 &  +40:34:33.91 &  7.99  &  2.20  &  00:39:48.25 &  +40:34:35.00 &  16.20  &  5.27  &  -0.46316 \\
  8  &  J003949+410421  &  AH0221-a  &  00:39:49.41 &  +41:04:21.22 &  7.80  &  2.71  &  00:39:49.41 &  +41:04:20.91 &  15.68  &  0.99  &  -0.45797 \\
  9  &  J003950+402657  &  AM0464-a  &  00:39:50.41 &  +40:26:57.10 &  1.80  &  0.06  &  00:39:50.24 &  +40:26:55.81 &  6.82  &  0.76  &  -0.87287 \\
  10  &  J003956+411134 &  AB0437-b  &  00:39:56.37 &  +41:11:34.90 &  63.62  &  0.06  &  00:39:56.14 &  +41:11:37.89 &  106.35  &  7.11  &  -0.33668 \\
    \smallskip
  11  &  J003957+411358  &  AH0524-a  &  00:39:57.59 &  +41:13:58.00 &  2.33  &  0.20  &  00:39:57.38 &  +41:13:58.60 &  6.46  &  0.57  &  -0.66822 \\
  12  &  J004005+401605  &  AC0308-c  &  00:40:05.32 &  +40:16:05.10 &  3.21  &  0.06  &  00:40:5.09 &  +40:16:8.15 &  5.40  &  0.64  &  -0.34082 \\
  13  &  J004006+402148  &  AM0464-a  &  00:40:06.40 &  +40:21:48.40 &  8.49  &  1.12  &  00:40:6.32 &  +40:21:45.83 &  21.63  &  1.29  &  -0.61319 \\
  14  &  J004013+410839  &  AH0221-a  &  00:40:13.28 &  +41:08:39.05 &  35.05  &  29.89  &  00:40:13.29 &  +41:08:42.35 &  68.40  &  4.71  &  -0.43811 \\
  15  &  J004016+405824  &  AH0221-a  &  00:40:16.93 &  +40:58:24.36 &  21.58  &  7.55  &  00:40:16.79 &  +40:58:25.32 &  61.06  &  4.45  &  -0.68159 \\
    \smallskip
  16  &  J004017+395508  &  AC0308-c  &  00:40:17.24 &  +39:55:08.10 &  15.54  &  0.06  &  00:40:17.13 &  +39:55:4.17 &  190.41  &  9.23  &  -1.64195 \\
  17  &  J004024+412926  &  AC0308-c  &  00:40:24.13 &  +41:29:26.70 &  6.09  &  0.06  &  00:40:24.11 &  +41:29:27.98 &  24.36  &  1.41  &  -0.90840 \\
  18  &  J004024+410712  &  AH0221-a  &  00:40:24.36 &  +41:07:12.95 &  26.96  &  11.04  &  00:40:24.51 &  +41:07:12.58 &  83.05  &  5.94  &  -0.73731 \\
  19  &  J004024+412029  &  AH0524-a  &  00:40:24.84 &  +41:20:29.50 &  8.82  &  0.55  &  00:40:24.65 &  +41:20:32.44 &  9.73  &  0.72  &  -0.06434 \\
  20  &  J004030+402755  &  AM0464-a  &  00:40:30.30 &  +40:27:55.50 &  6.18  &  1.55  &  00:40:30.29 &  +40:27:54.50 &  10.15  &  0.79  &  -0.32476 \\
   \smallskip
  21  &  J004035+413511  &  AC0308-c  &  00:40:35.58 &  +41:35:11.10 &  16.02  &  0.06  &  00:40:35.75 &  +41:35:10.89 &  12.81  &  0.82  &  0.14653 \\
  22  &  J004044+404846  &  AH0221-a  &  00:40:44.29 &  +40:48:46.59 &  12.66  &  4.15  &  00:40:44.35 &  +40:48:45.42 &  34.27  &  3.65  &  -0.65274 \\
  23  &  J004047+405525  &  AH0221-a  &  00:40:47.11 &  +40:55:25.17 &  2.60  &  0.43  &  00:40:47.05 &  +40:55:24.20 &  3.69  &  0.42  &  -0.22942 \\
  24  &  J004055+405723  &  AH0221-a  &  00:40:55.88 &  +40:57:23.48 &  28.16  &  2.61  &  00:40:55.90 &  +40:57:23.34 &  153.25  &  7.59  &  -1.11024 \\
  25  &  J004057+415438  &  AC0496-a  &  00:40:57.96 &  +41:54:38.20 &  1.17  &  0.06  &  00:40:58.08 &  +41:54:37.59 &  23.45  &  2.62  &  -1.96442 \\
    \smallskip
  26  &  J004100+411354  &  AH0524-a  &  00:41:00.26 &  +41:13:54.40 &  5.74  &  2.83  &  00:41:0.18 &  +41:13:54.90 &  12.79  &  0.86  &  -0.52558 \\
  27  &  J004103+410430  &  AH0524-a  &  00:41:03.03 &  +41:04:30.30 &  2.68  &  0.21  &  00:41:2.91 &  +41:04:27.99 &  3.36  &  0.41  &  -0.14940 \\
  28  &  J004111+402411  &  AH0221-a  &  00:41:11.30 &  +40:24:11.01 &  11.55  &  1.02  &  00:41:11.33 &  +40:24:10.59 &  51.26  &  2.58  &  -0.97631 \\
  29  &  J004111+403328  &  AM0464-a  &  00:41:11.43 &  +40:33:28.70 &  5.37  &  0.49  &  00:41:11.74 &  +40:33:28.97 &  15.67  &  0.96  &  -0.70174 \\
  30  &  J004112+412454  &  AH0524-a  &  00:41:12.82 &  +41:24:54.38 &  9.01  &  0.06  &  00:41:12.88 &  +41:24:57.51 &  28.65  &  2.89  &  -0.75803 \\
\hline
\end{tabular}
\vskip.5cm
\end{landscape}

\centerline{\includegraphics[width=0.5\textwidth,angle=270]{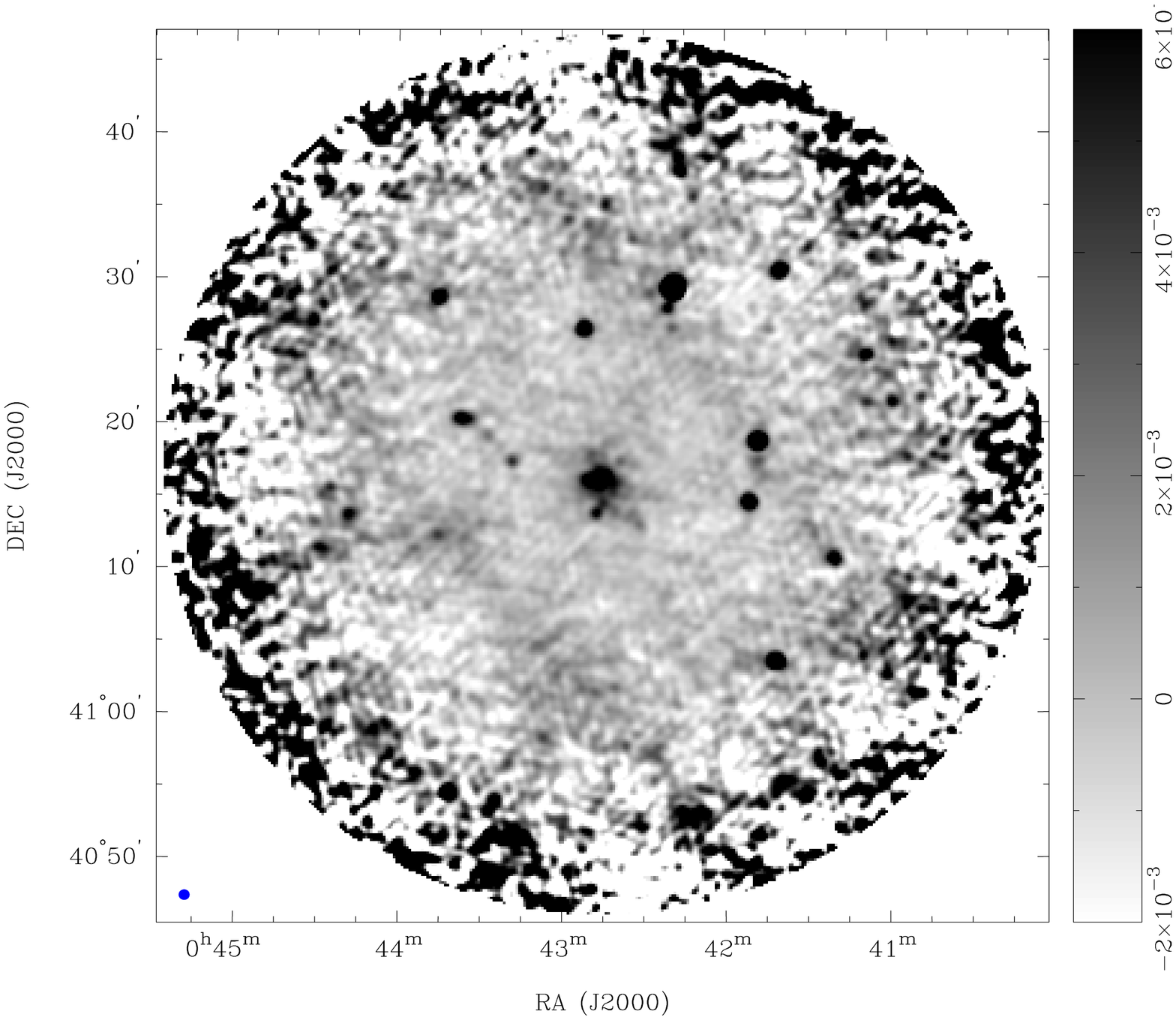}}
\figurecaption{1.}{VLA Project AC0101 radio-continuum total intensity image of \M. The synthesised beam, as represented by the blue circle in the lower left hand corner, is $45.9\arcsec \times 43.2$\arcsec\ and the r.m.s noise is 0.49~mJy/beam.}

\centerline{\includegraphics[width=0.5\textwidth,angle=270]{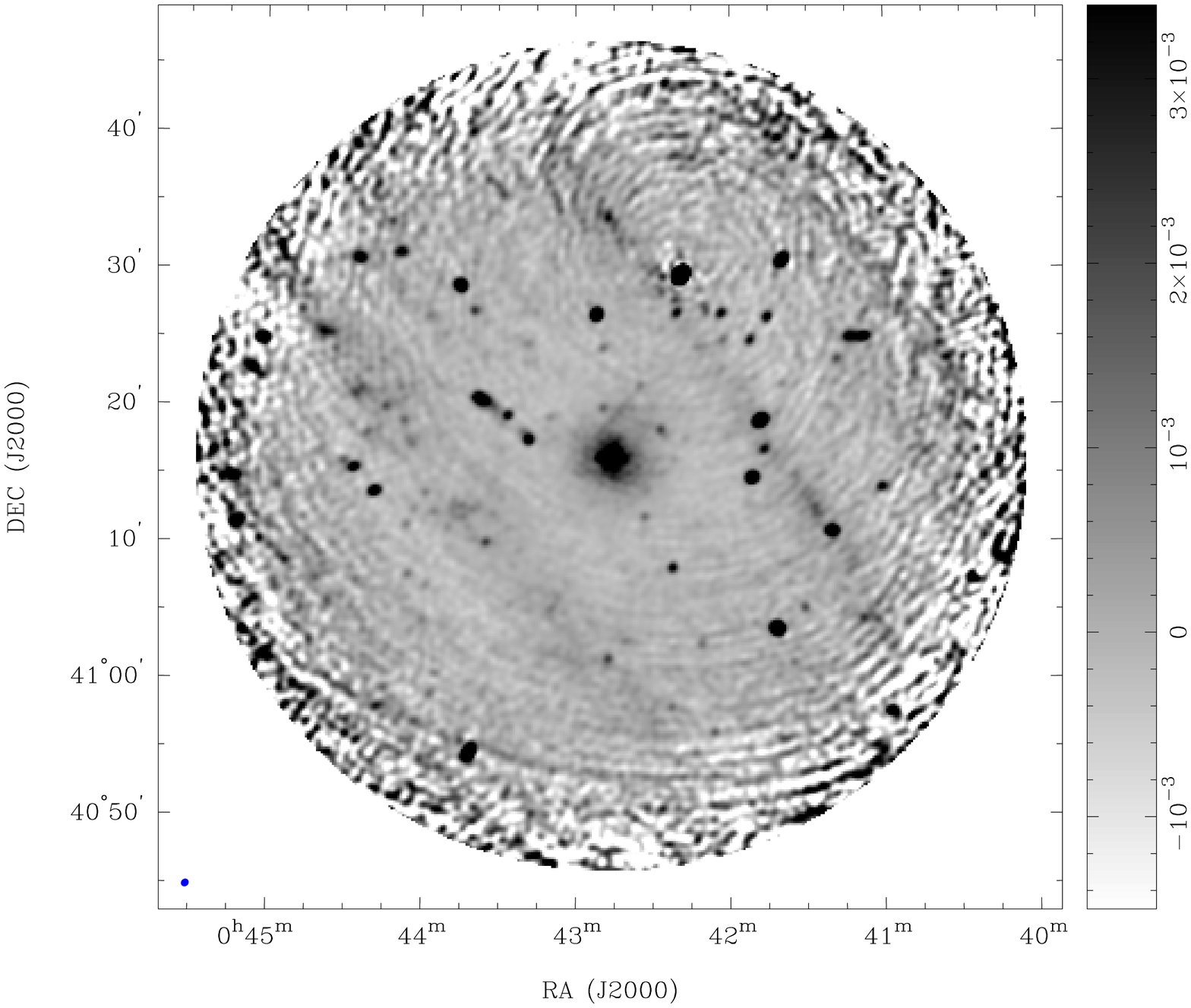}}
\figurecaption{2.}{VLA Project AB0551 radio-continuum total intensity image of \M. The synthesised beam, as represented by the blue circle in the lower left hand corner, is $35.9\arcsec \times 32.1$\arcsec\ and the r.m.s noise is 0.12~mJy/beam. This image is in terms of Jy/Beam. \label{fig:AB0551}}

\centerline{\includegraphics[width=0.5\textwidth,angle=270]{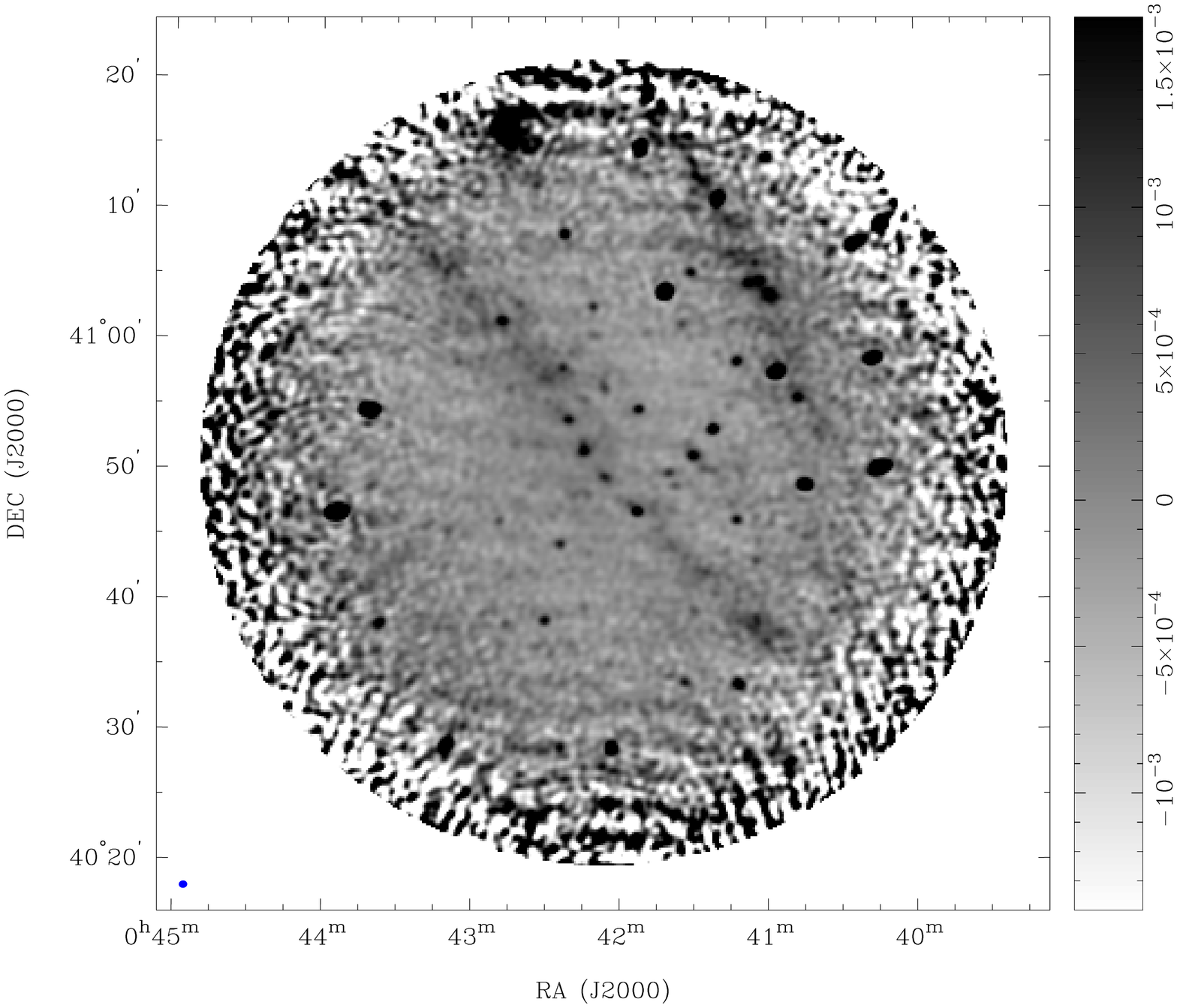}}
\figurecaption{3.}{VLA Project AB0491 radio-continuum total intensity image of \M. The synthesised beam, as represented by the blue circle in the lower left hand corner, is $39.0\arcsec \times 33.8$\arcsec\ and the r.m.s noise is 0.12~mJy/beam. This image is in terms of Jy/Beam. \label{fig:AB0491}}

\centerline{\includegraphics[width=0.5\textwidth,angle=270]{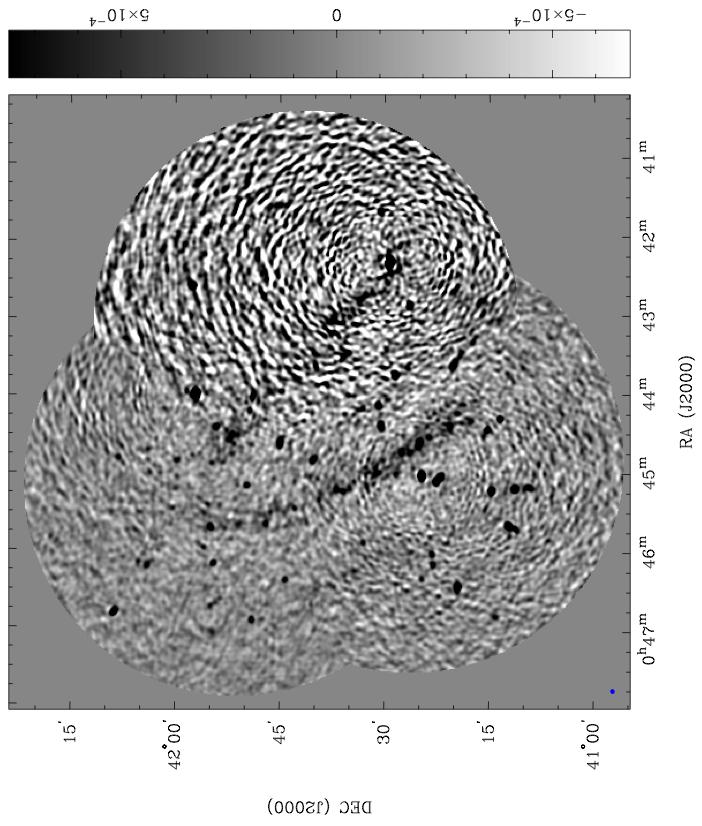}}
\figurecaption{4.}{VLA Project AB0647, segment a, radio-continuum total intensity image of \M. The synthesised beam, as represented by the blue circle in the lower left hand corner, is $41.2\arcsec \times 37.3$\arcsec\ and the r.m.s noise is 0.24~mJy/beam. This image is in terms of Jy/Beam. \label{fig:AB0647-a}}

\centerline{\includegraphics[width=0.5\textwidth,angle=270]{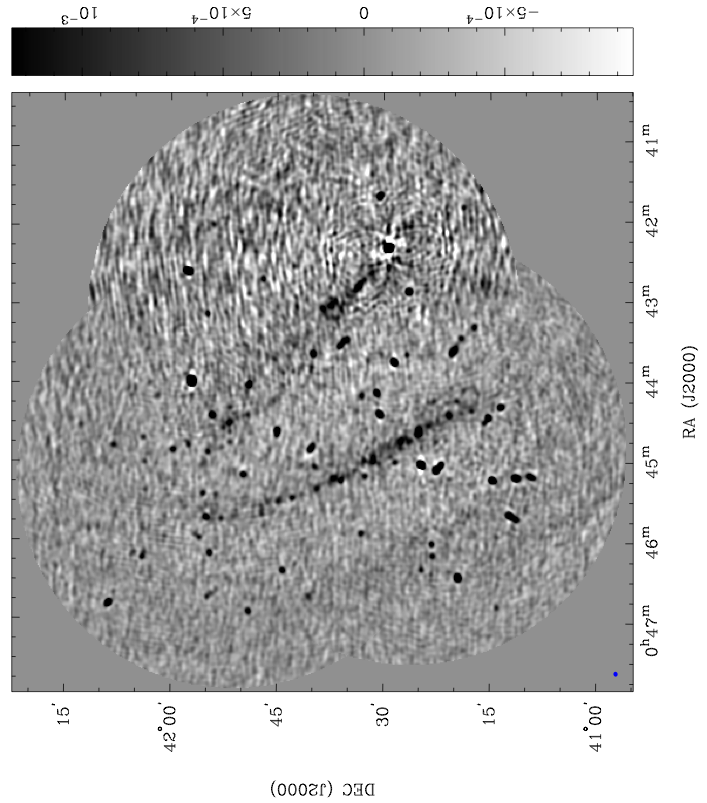}}
\figurecaption{5.}{VLA Project AB0647, segment b, radio-continuum total intensity image of \M. The synthesised beam, as represented by the blue circle in the lower left hand corner, is $40.9\arcsec \times 35.4$\arcsec\ and the r.m.s noise is 0.24~mJy/beam. This image is in terms of Jy/Beam. \label{fig:AB0647-b}}

\centerline{\includegraphics[width=0.5\textwidth,angle=270]{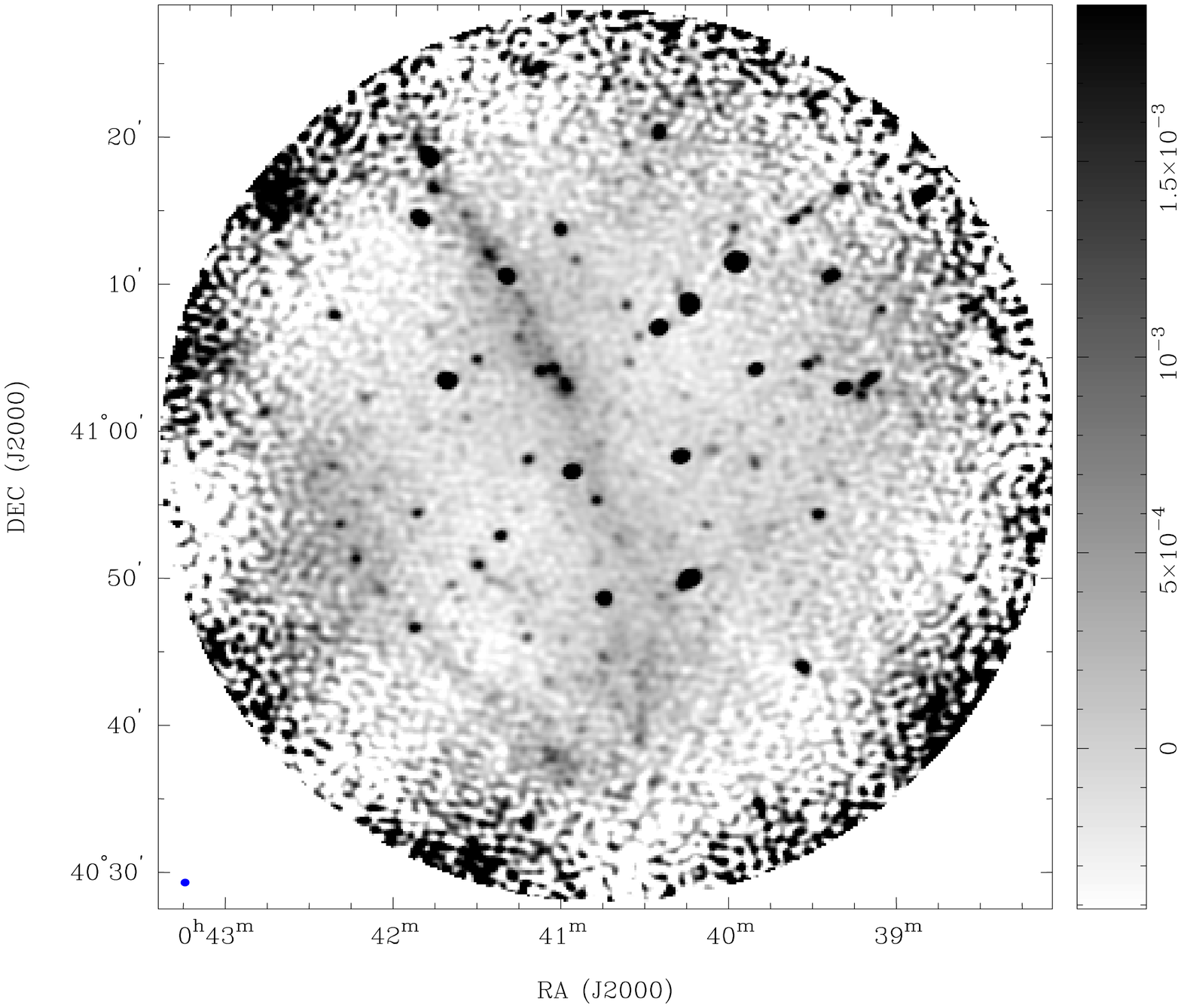}}
\figurecaption{6.}{VLA Project AB0437, segment a, radio-continuum total intensity image of \M. The synthesised beam, as represented by the blue circle in the lower left hand corner, is $36.0\arcsec \times 31.0$\arcsec\ and the r.m.s noise is 0.10~mJy/beam. This image is in terms of Jy/Beam. \label{fig:AB0437-a}}

\centerline{\includegraphics[width=0.5\textwidth,angle=270]{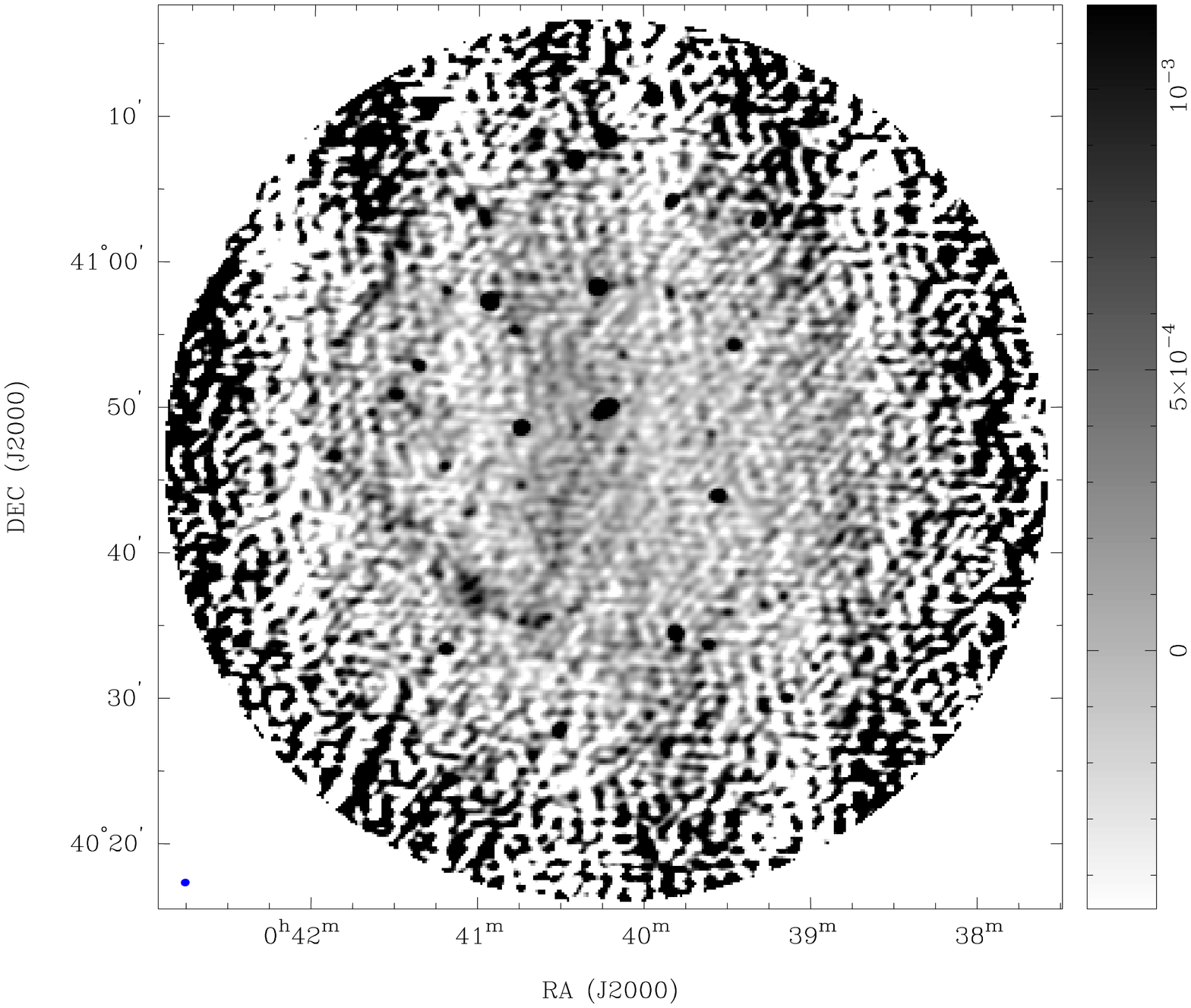}}
\figurecaption{7.}{VLA Project AB0437, segment b, radio-continuum total intensity image of \M. The synthesised beam, as represented by the blue circle in the lower left hand corner, is $36.0\arcsec \times 31.1$\arcsec\ and the r.m.s noise is 0.19~mJy/beam. This image is in terms of Jy/Beam. \label{fig:AB0437-b}}

\centerline{\includegraphics[width=0.5\textwidth,angle=270]{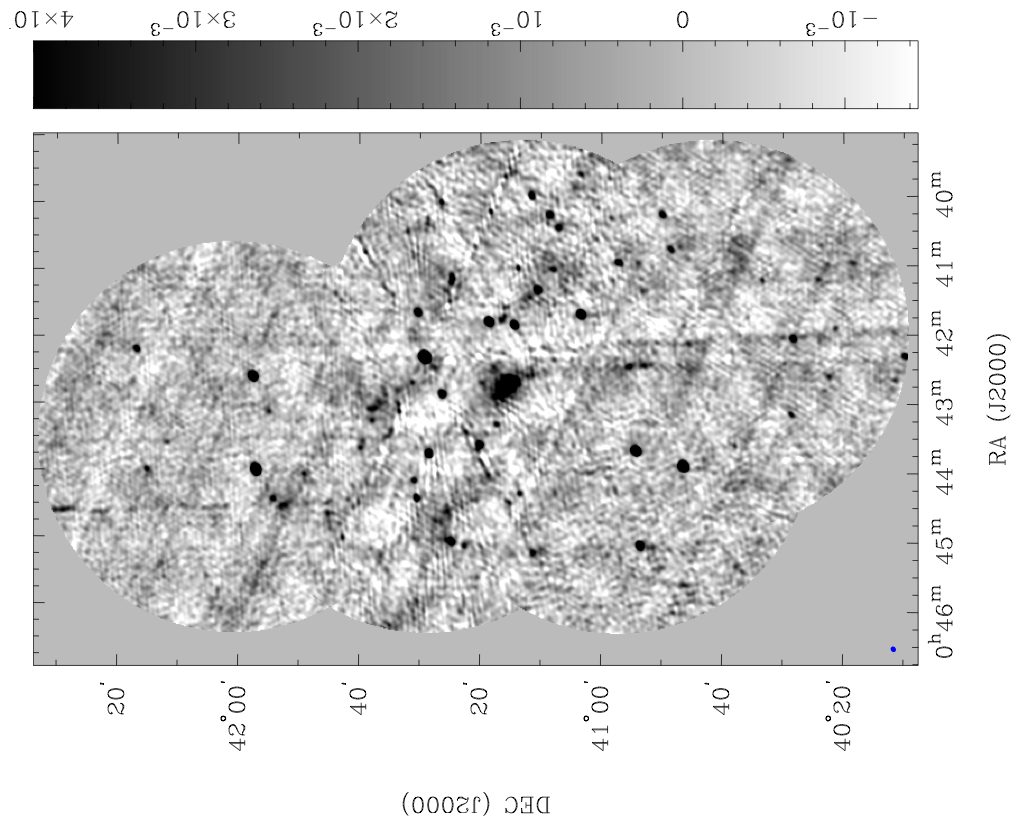}}
\figurecaption{8.}{VLA Project AC0308, segment a, radio-continuum total intensity image of \M. The synthesised beam, as represented by the blue circle in the lower left hand corner, is $57.9\arcsec \times 49.8$\arcsec\ and the r.m.s noise is 0.72~mJy/beam. This image is in terms of Jy/Beam. \label{fig:AC0308-a}}

\centerline{\includegraphics[width=0.5\textwidth,angle=270]{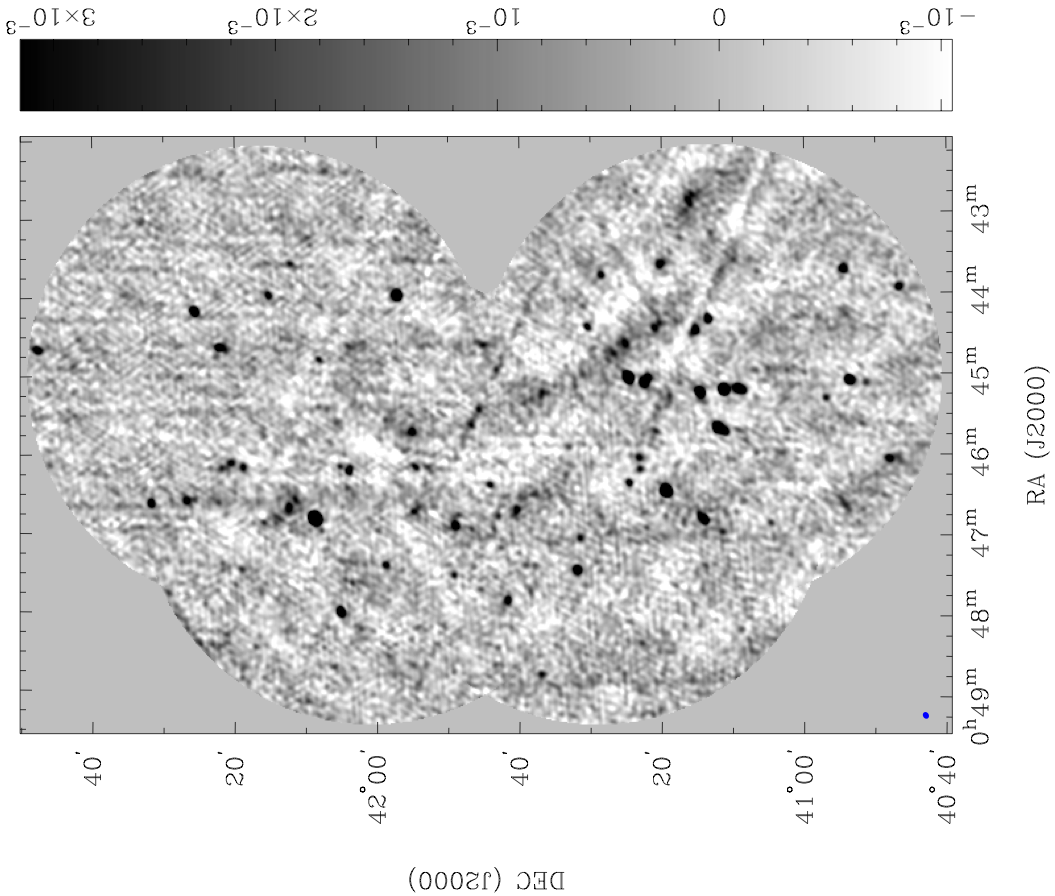}}
\figurecaption{9.}{VLA Project AC0308, segment b, radio-continuum total intensity image of \M. The synthesised beam, as represented by the blue circle in the lower left hand corner, is $58.9\arcsec \times 48.9$\arcsec\ and the r.m.s noise is 0.54~mJy/beam. This image is in terms of Jy/Beam. \label{fig:AC0308-b}}

\centerline{\includegraphics[width=0.5\textwidth,angle=270]{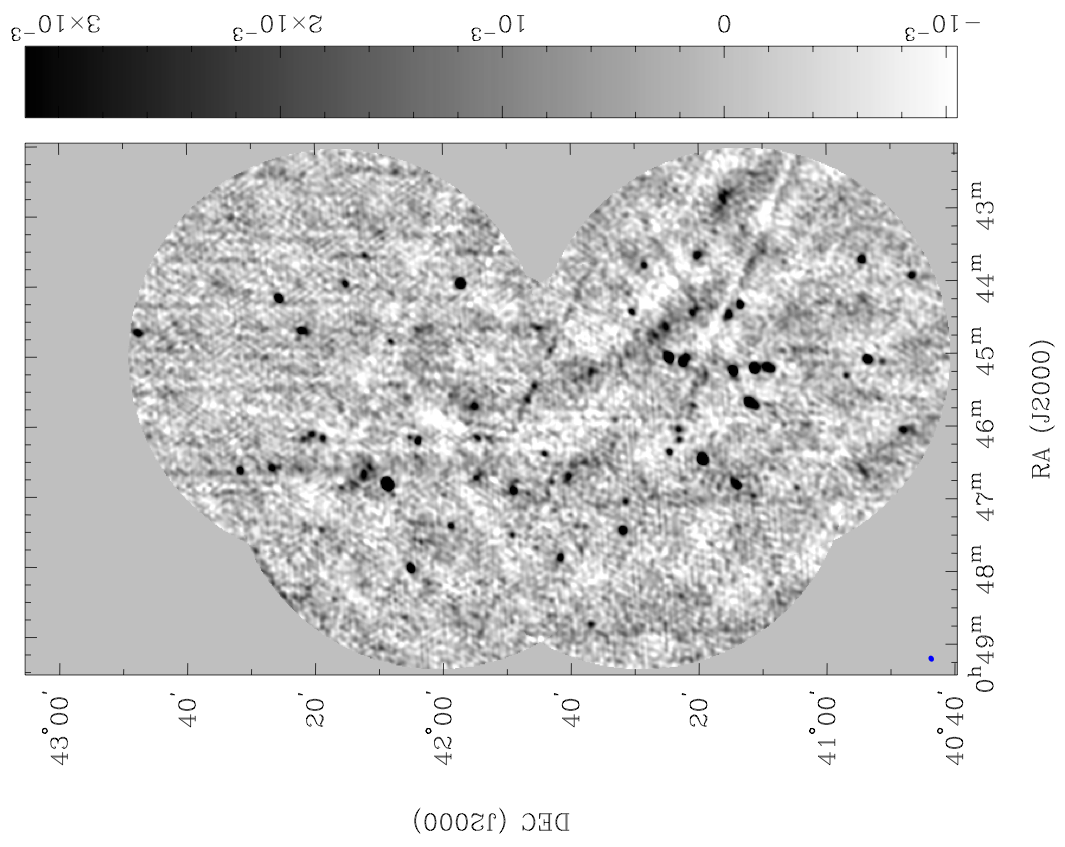}}
\figurecaption{10.}{VLA Project AC0308, segment c, radio-continuum total intensity image of \M. The synthesised beam, as represented by the blue circle in the lower left hand corner, is $58.0\arcsec \times 50.2$\arcsec\ and the r.m.s noise is 0.60~mJy/beam. This image is in terms of Jy/Beam. \label{fig:AC0308-c}}

\centerline{\includegraphics[width=0.5\textwidth,angle=270]{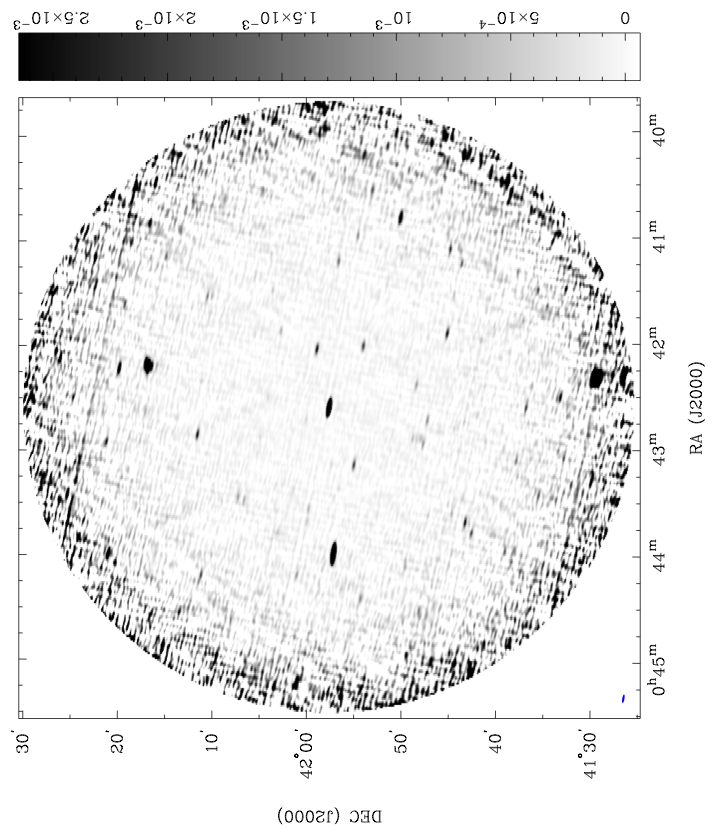}}
\figurecaption{11.}{VLA Project AC0496 radio-continuum total intensity image of \M. The synthesised beam, as represented by the blue circle in the lower left hand corner, is $54.6\arcsec \times 14.0$\arcsec\ and the r.m.s noise is 0.08~mJy/beam. This image is in terms of Jy/Beam. \label{fig:AC0496}}

\centerline{\includegraphics[width=0.5\textwidth,angle=270]{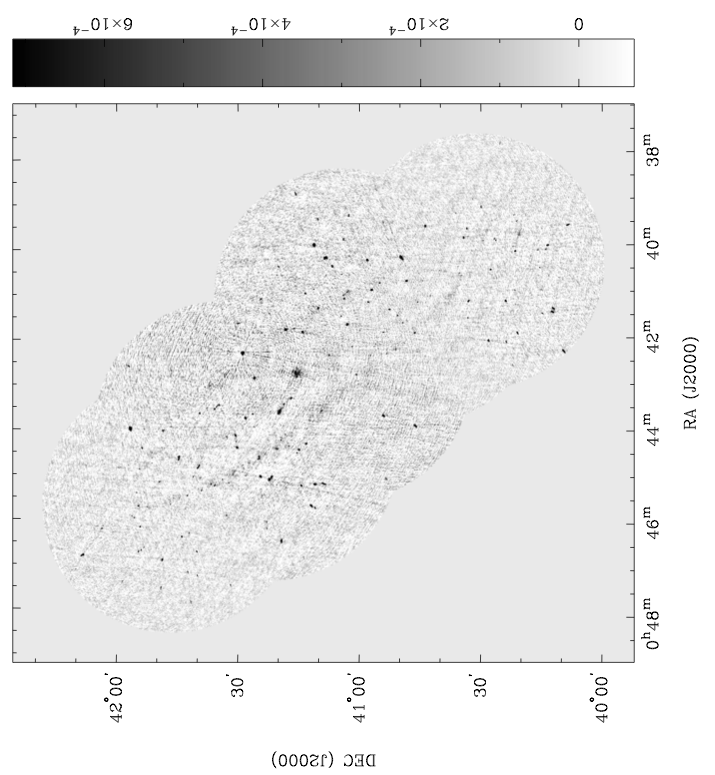}}
\figurecaption{12.}{VLA Project AM0464 radio-continuum total intensity image of \M. The synthesised beam is $12.8\arcsec \times 12.2$\arcsec\ and the r.m.s noise is 0.13~mJy/beam. This image is in terms of Jy/Beam. \label{fig:AM0464}}

\centerline{\includegraphics[width=0.5\textwidth,angle=270]{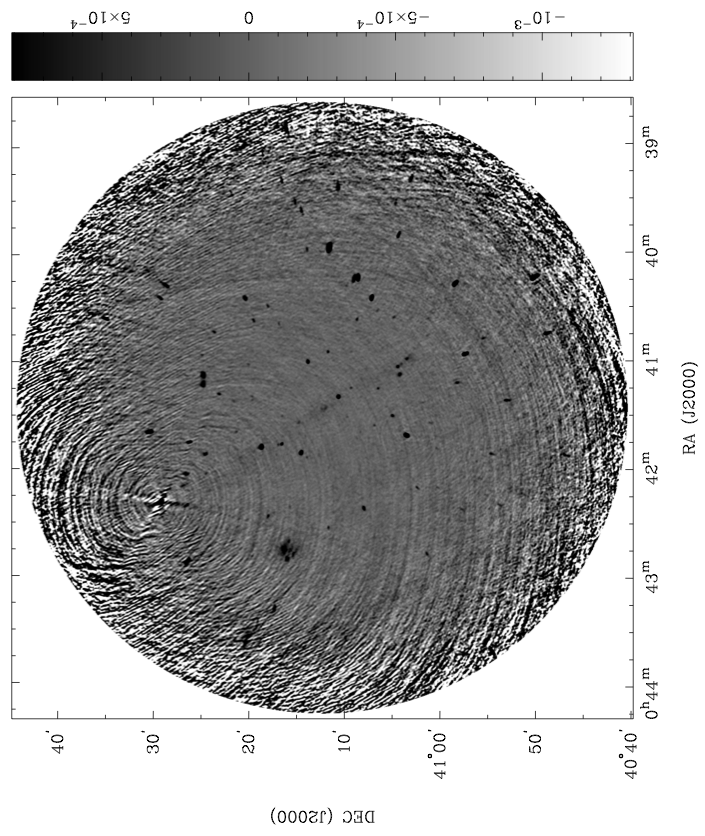}}
\figurecaption{13.}{VLA Project AH0524 radio-continuum total intensity image of \M. The synthesised beam is $12.8\arcsec \times 12.2$\arcsec\ and the r.m.s noise is 0.07~mJy/beam. This image is in terms of Jy/Beam. \label{fig:AH0524}}

\centerline{\includegraphics[width=0.5\textwidth,angle=270]{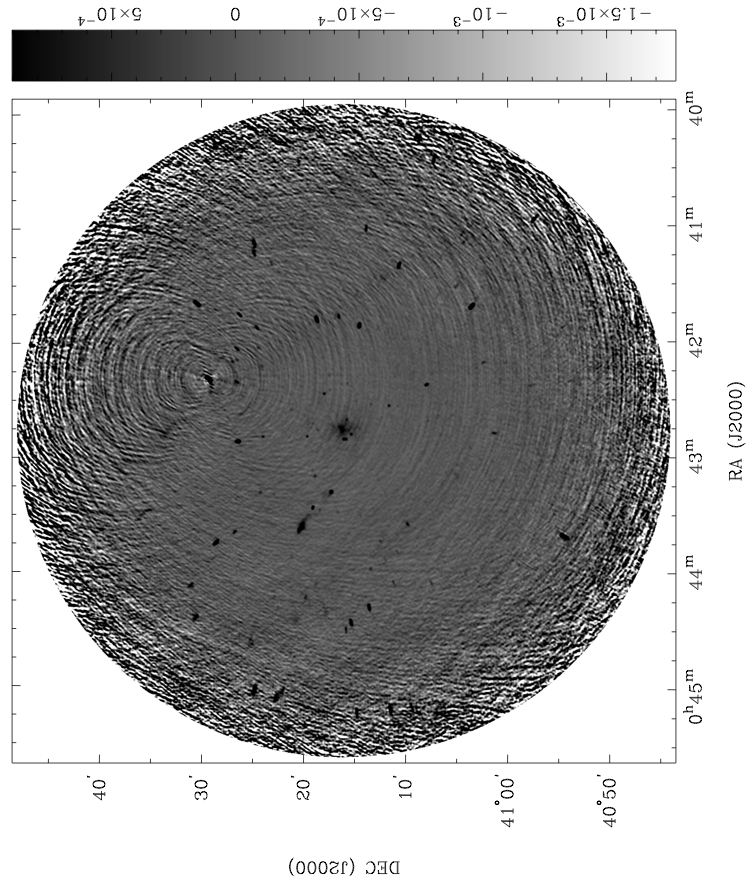}}
\figurecaption{14.}{VLA Project AB0679, segment a, radio-continuum total intensity image of \M. The synthesised beam is $12.0\arcsec \times 11.7$\arcsec\ and the r.m.s noise is 0.07~mJy/beam. This image is in terms of Jy/Beam. \label{fig:AB0679-a}}

\centerline{\includegraphics[width=0.5\textwidth,angle=270]{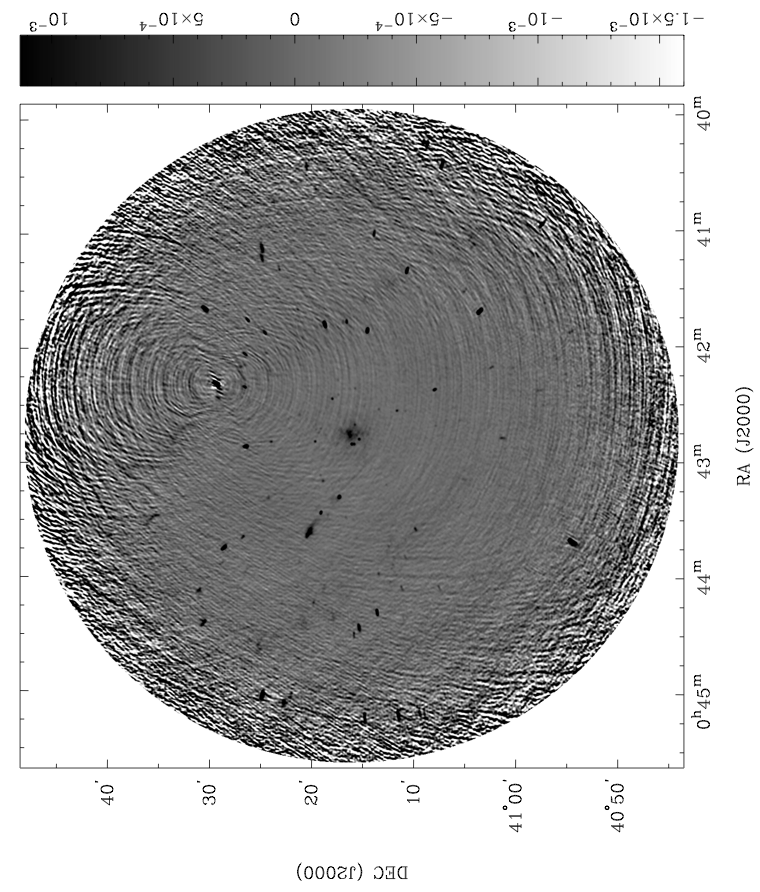}}
\figurecaption{15.}{VLA Project AB0679, segment b, radio-continuum total intensity image of \M. The synthesised beam is $12.1\arcsec \times 11.5$\arcsec\ and the r.m.s noise is 0.08~mJy/beam. This image is in terms of Jy/Beam. \label{fig:AB0679-b}}

\centerline{\includegraphics[width=0.5\textwidth,angle=270]{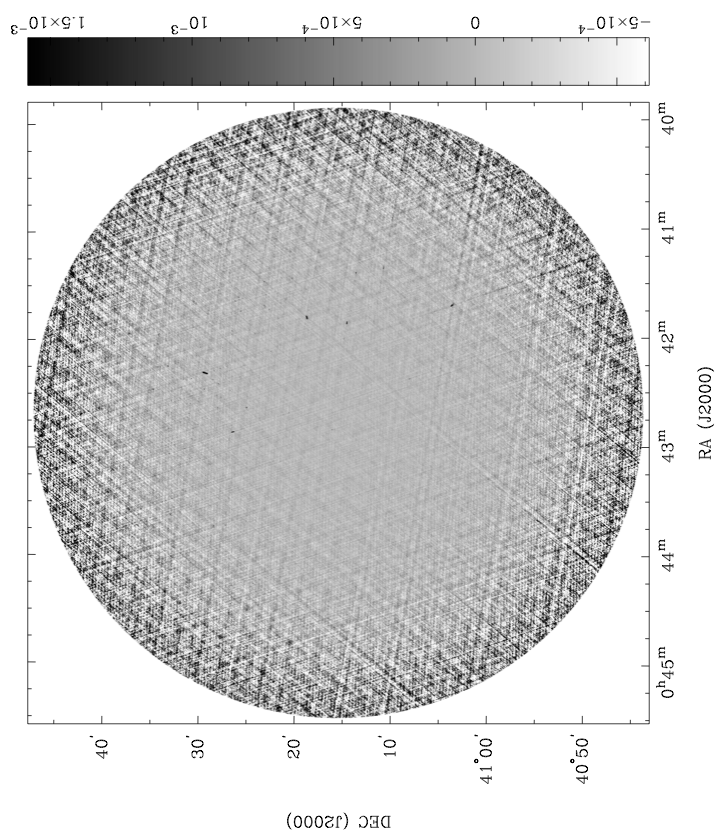}}
\figurecaption{16.}{VLA Project AT0149 radio-continuum total intensity image of \M. The synthesised beam is $4.0\arcsec \times 3.4$\arcsec\ and the r.m.s noise is 0.08~mJy/beam. This image is in terms of Jy/Beam. \label{fig:AT0149}}

\centerline{\includegraphics[width=0.5\textwidth,angle=270]{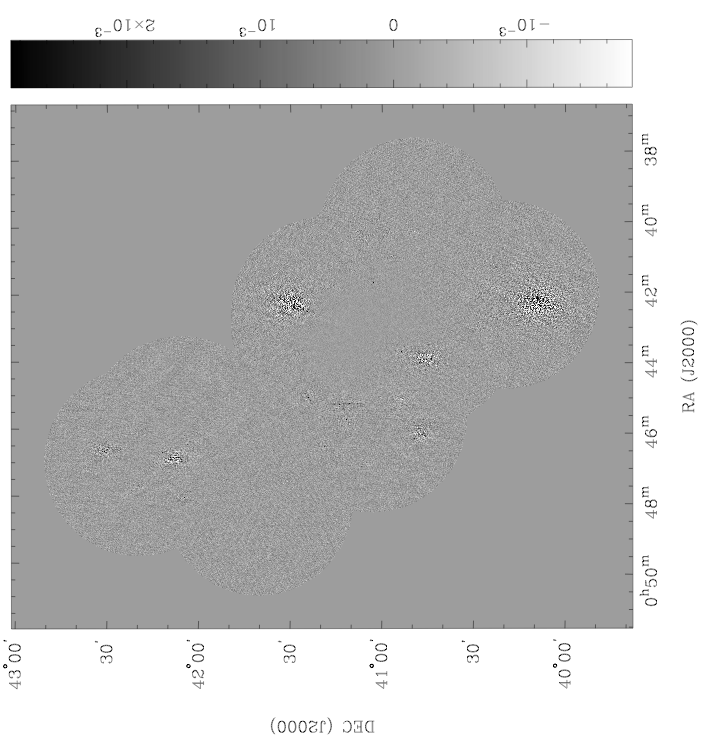}}
\figurecaption{17.}{VLA Project AH0221 radio-continuum total intensity image of \M. The synthesised beam is $3.4\arcsec \times 3.2$\arcsec\ and the r.m.s noise is 0.22~mJy/beam. This image is in terms of Jy/Beam. \label{fig:AH0221}}

\centerline{\includegraphics[width=0.5\textwidth,angle=270]{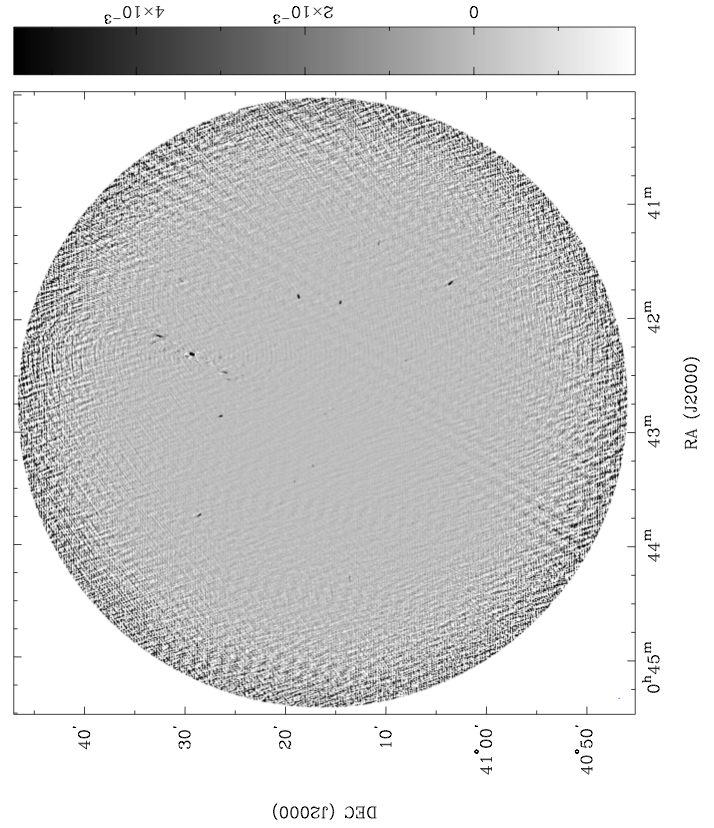}}
\figurecaption{18.}{VLA Project AH0139 radio-continuum total intensity image of \M. The synthesised beam is $7.2\arcsec \times 6.6$\arcsec\ and the r.m.s noise is 0.16~mJy/beam. This image is in terms of Jy/Beam. \label{fig:AH0139}}

\centerline{\includegraphics[width=\textwidth,trim=0 90 0 0,angle=270]{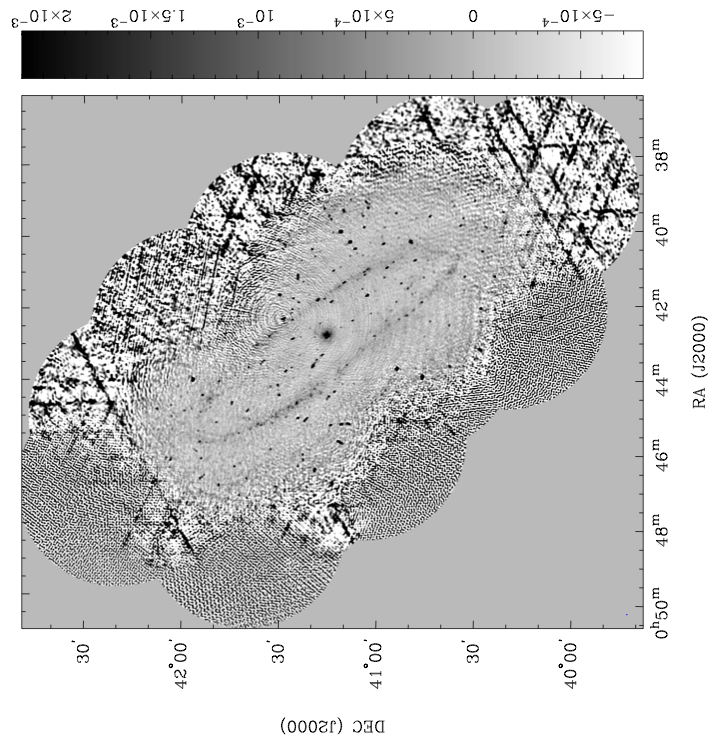}}
\figurecaption{19.}{A mosiac radio-continuum total intensity image of \M produced with all fully polarised VLA observations with a uv restriction of 0-5~k$\lambda$. The synthesised beam, as represented by the blue circle in the lower left hand corner, is $35.73\arcsec \times 16.38$\arcsec\ and the r.m.s noise is 0.145~mJy/beam. This image is in Jy/Beam. \label{fig:M31-all-uv5}}

\centerline{\includegraphics[width=\textwidth,trim=0 90 0 0,angle=270]{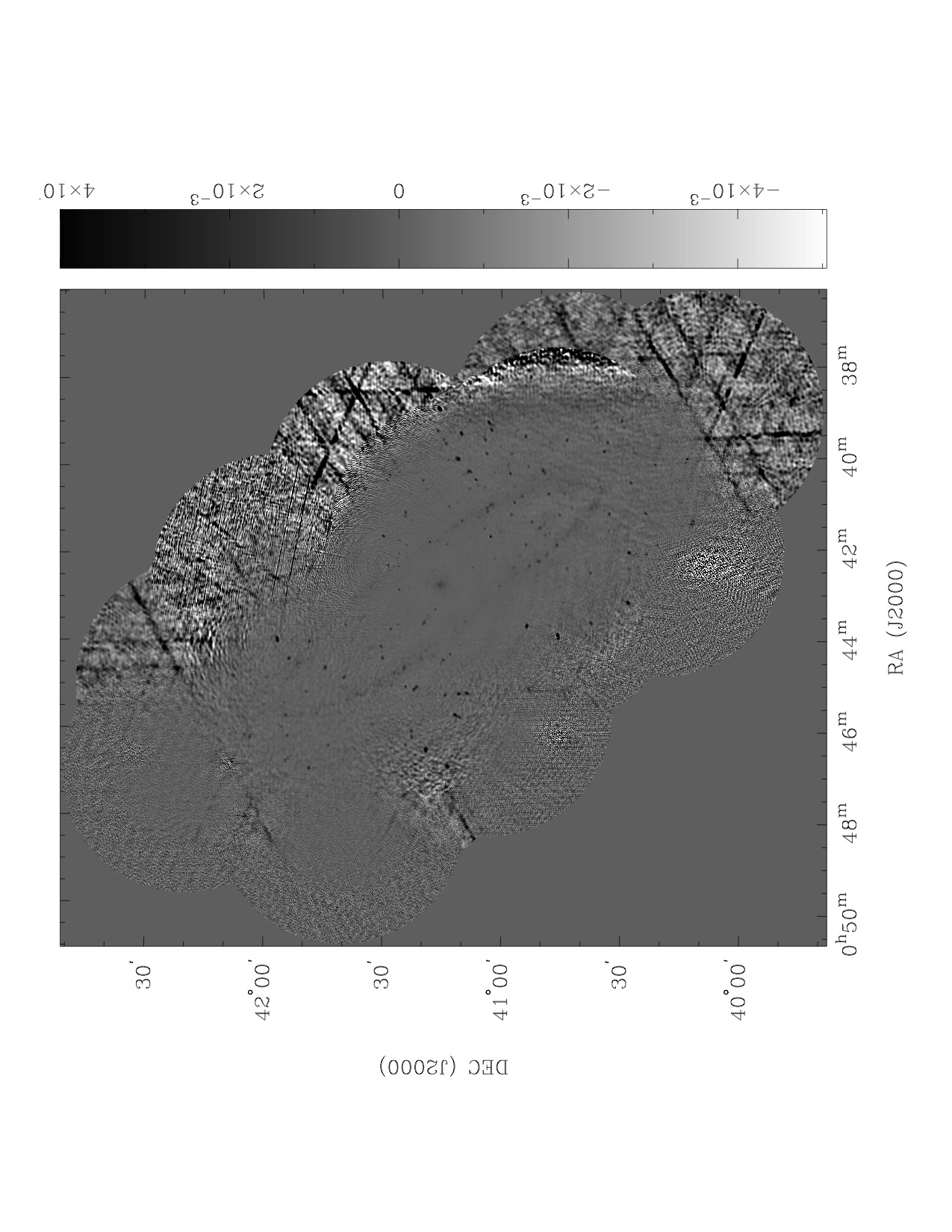}}
\figurecaption{20.}{A mosiac radio-continuum total intensity image of \M produced with all fully polarised VLA observations. The synthesised beam is $6.14\arcsec \times 5.35$\arcsec\ and the r.m.s noise is 0.09~mJy/beam. This image is in Jy/Beam. \label{fig:M31-all-uv25}}

\centerline{\includegraphics[width=\textwidth,trim=0 90 0 0,angle=270]{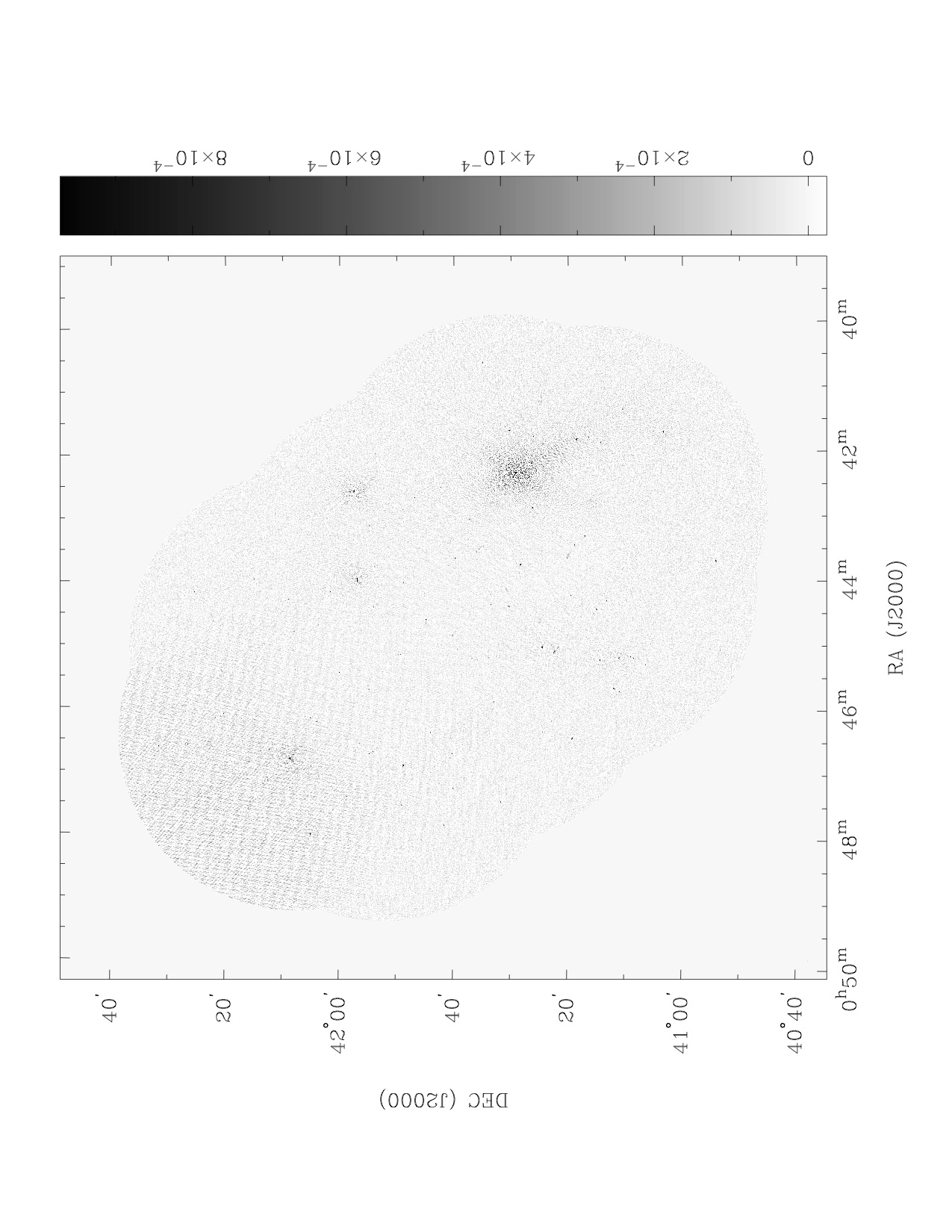}}
\figurecaption{21.}{A mosiac radio-continuum total intensity image of \M produced with VLA  project AB0396 and AB0999. The synthesised beam is $4.63\arcsec \times 3.78$\arcsec\ and the r.m.s noise is 0.08~mJy/beam. This image is in terms of Jy/Beam. \label{fig:M31-RCP-uv35}}

\centerline{\includegraphics[width=\textwidth,trim=0 90 0 0,angle=270]{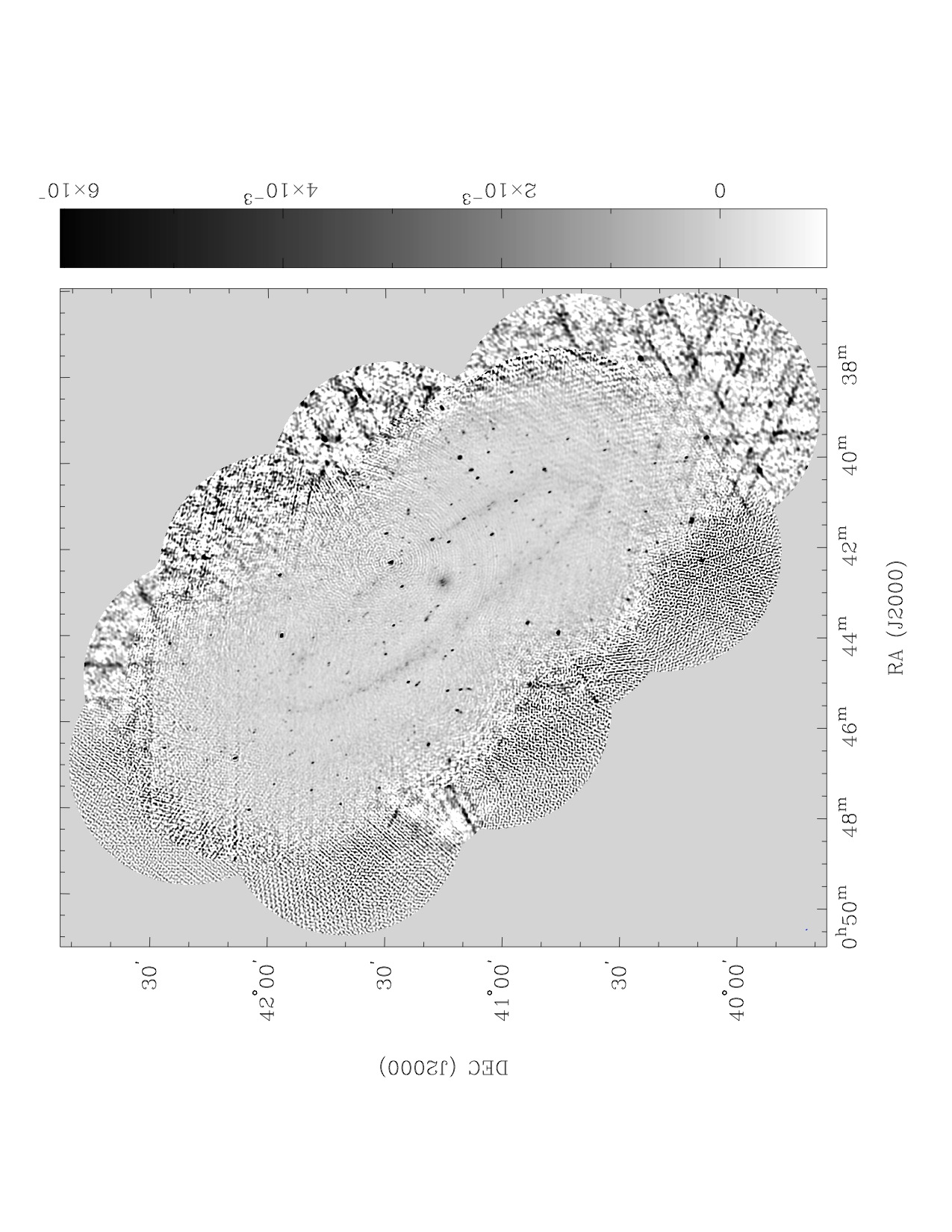}}
\figurecaption{22.}{A mosiac radio-continuum total intensity image of \M produced with all calibrated VLA observations. The synthesised beam, as represented by the blue circle in the lower left hand corner, is $32.61\arcsec \times 16.36$\arcsec\ and the r.m.s noise is 0.13~mJy/beam. This image is in terms of Jy/Beam. \label{fig:M31-uv5-allRCP}}

\centerline{\includegraphics[width=\textwidth,trim=0 90 0 0,angle=270]{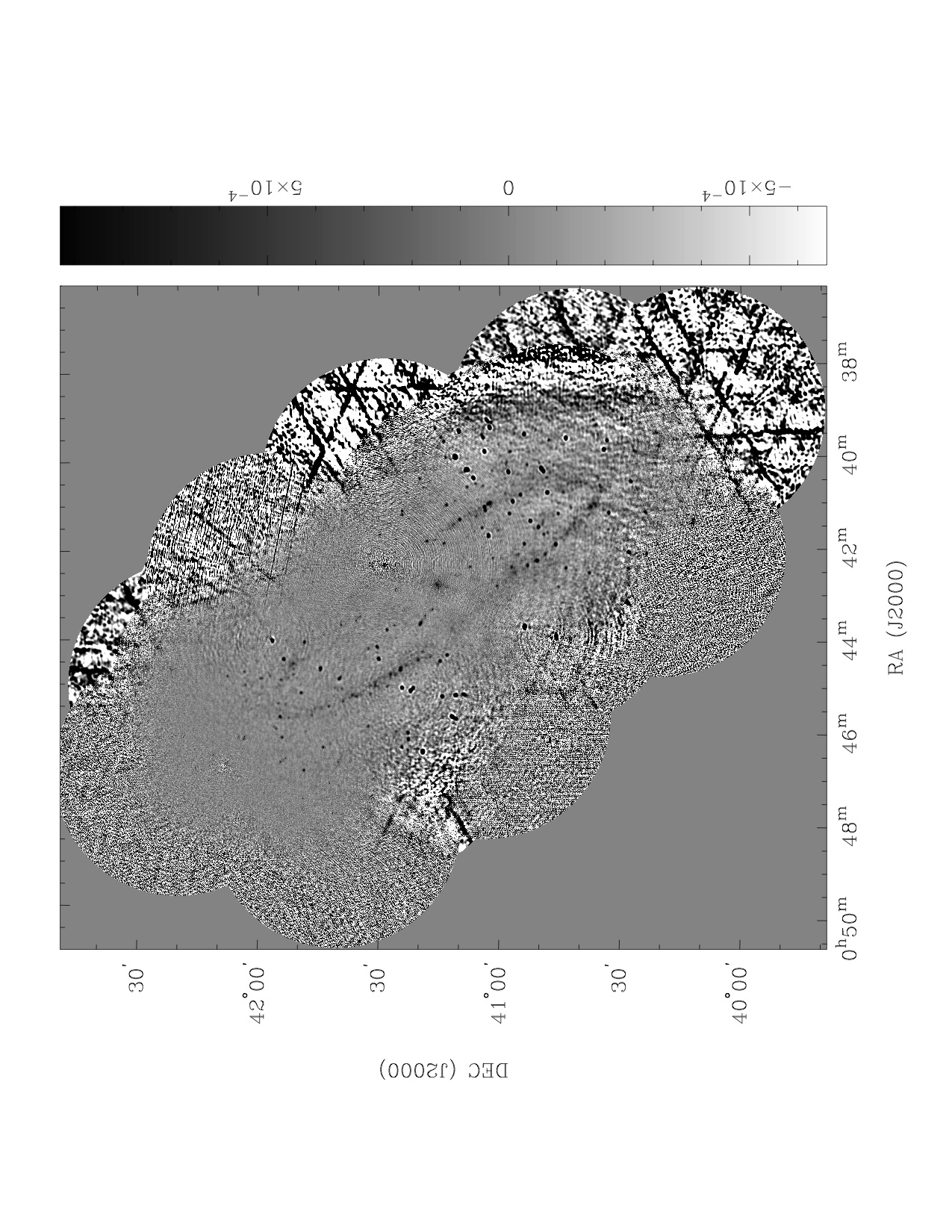}}
\figurecaption{23.}{A mosiac radio-continuum total intensity image of \M produced with all calibrated VLA observations. The synthesised beam is $6.13\arcsec \times 5.35$\arcsec\ and the r.m.s noise is 0.12~mJy/beam. This image is in terms of Jy/Beam. \label{fig:M31-uv25-allRCP}}

\centerline{\includegraphics[width=\textwidth,trim=150 40 100 40]{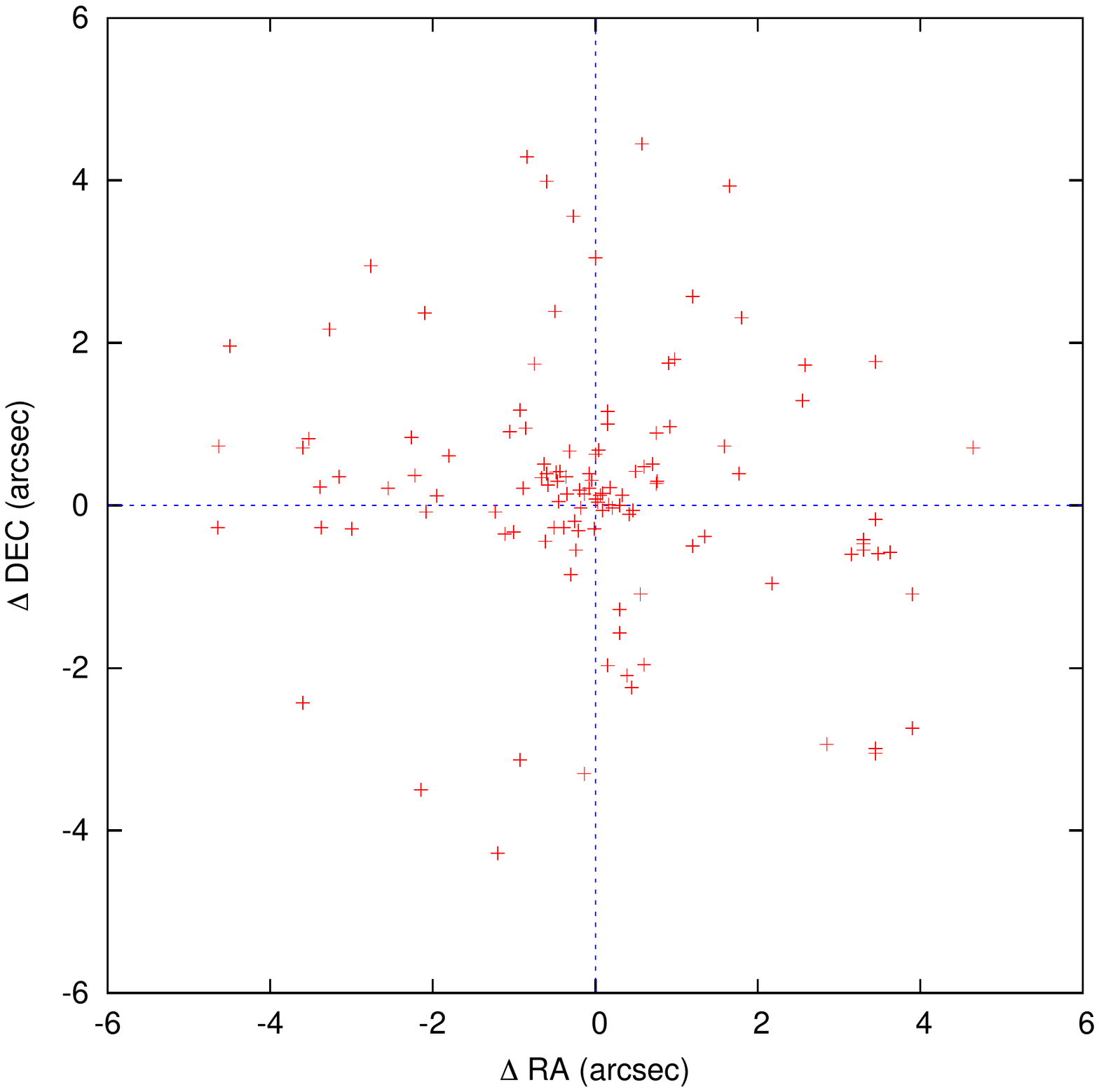}}
\figurecaption{24.}{Comparison of positional differences ($\Delta$RA and $\Delta$DEC) between our catalogue and Gelfand et al. (2004).\label{fig:GelfandUs}}

\centerline{\includegraphics[width=0.9\textwidth, trim=2cm 4cm 0cm 0cm]{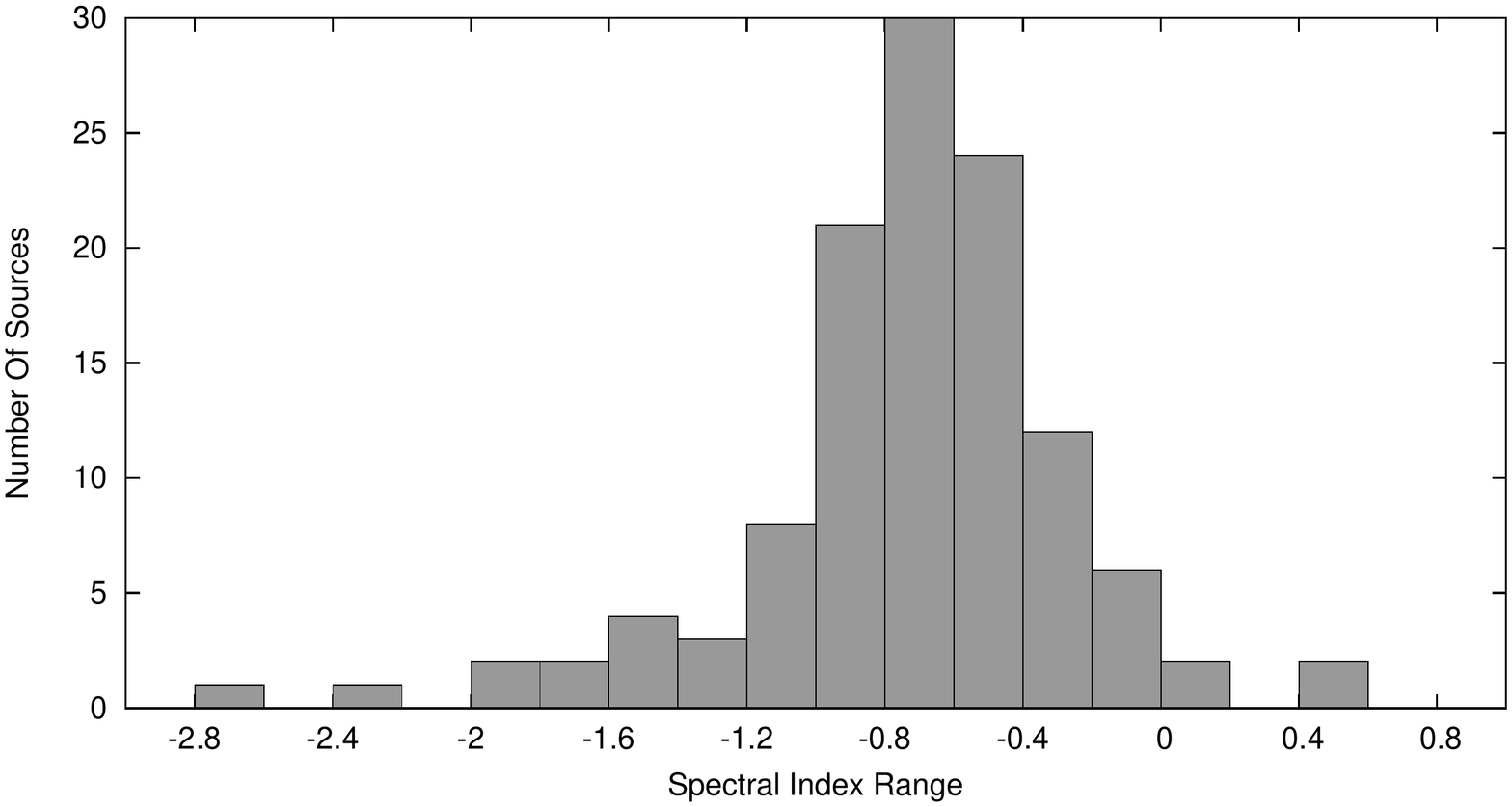}}
\figurecaption{25.}{Spectral index distribution of point sources in the field of \M.\label{fig:spectralhist}}

\begin{multicols}{2}
{

\section{6. Conclusions}

We present new $\lambda$=20~cm ($\nu$=1.4~GHz) images of \M\ constructed from archived VLA radio-continuum observations. These new images consisting of 17 individual VLA projects which are of high-sensitivity and resolution. Images presented here are sensitive to rms=60~$\mu$Jy and feature a high angular resolution ($<$10\arcsec). Also, we present a complete sample of 864 unique discrete radio sources across the field of \M. The most prominent region in \M\ is ``the ring feature'' for which we estimate a total integrated flux density of 706~mJy at $\lambda$=20~cm. From our 20-cm catalogue, we find 118 discrete sources that are in common to those listed in Gelfand et al. (2004) at $\lambda$=92~cm. The majority (61\%) of these sources exhibit a spectral index of $\alpha$$<$--0.6 indicating predominant non-thermal emission which is more typical of background objects.


\acknowledgements{We used the {\sc karma} and {\sc miriad} software packages developed by the ATNF. The National Radio Astronomy Observatory is a facility of the National Science Foundation operated under cooperative agreement by Associated Universities, Inc. }


\references

Braun, R.: 1990a, \journal{Aust. J. Phys. Astrophys. Suppl.}, \vol{72}, 755.

Braun, R.: 1990b, \journal{Aust. J. Phys. Astrophys. Suppl.}, \vol{72}, 761.

Dickel, J. R., Dodorico, S.: 1984, \journal{Mon. Not. R. Astron. Soc.}, \vol{206}, 351.

Dickel, J. R., Dodorico, S., Felli, M., Dopita, M.: 1982, \journal{Aust. J. of Phys.}, \vol{252}, 582.

Galvin, T. J.: et al., 2012, \journal{Astrophys. Space Sci.}, in press.

Gelfand, J. D., Lazio, T. J. W., Gaensler, B. M., 2004, \journal{Aust. J. Phys. Astrophys. Suppl.}, \vol{155}, 89.

Gooch, R.: 1996, in "Astronomical Society of the Pacific Conference Series, Vol. 101, Astronomical Data Analysis Software and Systems V", G. H. Jacoby \& J. Barnes, ed., 80.

Karachentsev, I. D., Karachentseva, V. E., Huchtmeier, W. K., Makarov, D. I.:  2004, \journal{Astron. J.}, \vol{127}, 2031.

Payne, J. L., Filipovi{\'c}, M. D., Pannuti, T. G., Jones, P. A., Duric N., White, G. L., Carpano, S.: 2004, \journal{Astron. Astrophys.}, \vol{425}, 443.

Sault, R. J., Teuben, P. J., Wright, M. C. H.: 1995, in "Astronomical Society of the Pacific Conference Series, Vol. 77, Astronomical Data Analysis Software and Systems IV", R. A. Shaw, H. E. Payne, \& J. J. E. Hayes, ed., 433.

Steer, D. G., Dewdney, P. E., Ito, M. R.: 1984, \journal{Astron. Astrophys.}, 137, 159.

\endreferences

}
\end{multicols}

\vskip.5cm

\vfill\eject

{\ }




\naslov{NOVO PROUQAVA{NJ}E M\,31 U RADIO-KONTINUMU NA 20~CM -- MAPE I KATALOZI TAQKASTIH IZVORA}


\authors{T.~J. Galvin, M. D. Filipovi\'c, E. J. Crawford,  N.~F.~H. Tothill, G. F. Wong, A.~Y. De~Horta}

\vskip3mm


\address{University of Western Sydney, Locked Bag 1797, Penrith South DC, NSW 2751, AUSTRALIA}

\Email{m.filipovic}{uws.edu.au}

\vskip3mm


\centerline{\rrm UDK \udc}

\vskip1mm

\centerline{\rit Originalni nauqni rad}

\vskip.7cm

\begin{multicols}{2}

{


\rrm

U ovoj studiji predstav{lj}amo nove {\rm Very Large Array (VLA)} radio-kontinum mape i kataloge taqkastih objekata u po{lj}u M\,31 na {\rm $\lambda$=20~cm ($\nu$=1.4~GHz)}. Nove mape visoke rezolucije {\rm ($<$10\arcsec)} i oset{lj}ivosti {\rm (rms=60~$\mu$Jy)} su naprav{lj}ene spaja{nj}em svih 17 arhiviranih posmatra{nj}a {\rm VLA} teleskopa. Kompletan katalog svih objekata u po{lj}u M\,31 galaksije sadr{\zz}i 864 radio izvora. Ovi objekti su upore{dj}eni sa {\rm Gelfand et al. (2004)} katalogom na {\rm $\lambda$=92~cm} i na{dj}eno je 118 zajedniqih radio izvora u oba kataloga. Ve{\cc}ina ovih objekata (61\%) imaju veoma strm radio spektralni indeks {\rm ($\alpha <$--0.6)} xto je tipiqno za netermalne izvore koji se nalaze van M\,31 galaksije. Detektovali smo i jedan od najprominentnijih oblasti u M\,31 galaksiji -- prsten --  sa ukupnom gustinom fluksa od {\rm 706~mJy} na {\rm $\lambda$=20~cm.}

}

\end{multicols}

\end{document}